\RequirePackage{fix-cm}
\documentclass[twocolumn]{svjour3}          
\smartqed  
\usepackage[latin1]{inputenc}
\usepackage{graphicx}
\usepackage{bm}
\usepackage{multirow}
\usepackage{subfig}
\usepackage{caption}
\usepackage [round]{natbib}
	\setcitestyle{aysep={}} 
\usepackage{amsmath}
\usepackage{color} 

\usepackage[toc,page]{appendix}
\usepackage{makecell}
\usepackage{rotating}
\usepackage{multirow}
\journalname{Advances in Data Analysis and Classification}
\begin{document}

\title{Basis expansion approaches for Functional Analysis of Variance with repeated measures}


\author{Christian Acal         \and
	Ana M. Aguilera 
}


\institute{C. Acal, A.M. Aguilera \at
	Department of Statistics and Operations Research and IMAG, University of
	Granada, Granada, Spain \\
	\email{chracal@ugr.es}, \email{aaguiler@ugr.es}         
}

\date{Received: date / Accepted: date}

\maketitle

\begin{abstract}
The methodological contribution in this paper is motivated by biomechanical studies where data characterizing human movement are waveform curves representing joint measures such as flexion angles, velocity, acceleration, and so on. In many cases the aim consists of detecting differences in gait patterns when several independent samples of subjects walk or run under different conditions (repeated measures). Classic kinematic studies often analyse discrete summaries of the sample curves discarding important information and providing biased results.  As the sample data are obviously curves, a Functional Data Analysis approach is proposed to solve the problem of testing the equality of the mean curves of a functional variable observed on several independent groups under different treatments or time periods. A novel approach for Functional Analysis of Variance (FANOVA) for repeated measures that takes into account the complete curves is introduced.  By assuming a basis expansion for each sample curve, two-way FANOVA problem is reduced to Multivariate ANOVA for the multivariate response of basis coefficients. Then, two different approaches for MANOVA with repeated measures are considered. Besides, an extensive simulation study is developed to check their performance. Finally, two applications with gait data are developed.   \\

\keywords{Functional Data Analysis \and Multivariate Analysis of Variance \and Repeated Measures \and Splines \and Biomechanics}
\end{abstract}

\section{Introduction}

\sloppy 

The well known Analysis of Variance (ANOVA) methodology aims at comparing more than two groups and/or treatments with respect to an scalar response variable. This comparison is based on the total variability decomposition of an experiment in independent components that are attributed to different \textcolor{red}{reasons}. In general terms, it determines if the discrepancy between the averages of the treatments is greater than what would be expected within the treatments.  Thus, ANOVA can be seen as the natural extension of two-sample classical statistical tests  to the case of more than two populations. From its formulation,  ANOVA has been constantly object of study to adapt it for different scenarios: single or multiple factors, parametric-non parametric cases,   repeated measures frameworks or longitudinal data, among others. The more complex  problem of testing the equality of a large number of populations has been recently studied in \cite{jimenez2021b}. Motivated by biomechanical studies where the aim is to test the differences between the means of the human movement curves observed on independent  samples of subjects under different treatments, the current manuscript is focused on addressing this tool from a Functional Data Analysis (FDA) perspective.    

FDA is a branch of statistics devoted  to analyse the information of functions (usually curves) that evolve over time, space or another continuous argument. FDA is able to explore the curves in all the domain. This fact avoids the loss of important information  usually produced when functional data are analysed through multivariate approaches from discrete measures.  The great computational advances experimented by technology  in last years make possible to model the complete curve and  retain its  main features.  Key text books in FDA and related topics offer a broad vision of the most general methodologies, applications and computational aspects of this field, see e.g. \cite{Ramsay02,Ramsay05,Ramsay09,FerratyVieu,Horvath}.  The majority of the classical multivariate statistical techniques have been extended for functional data:  principal and independent component analysis \citep{Aguilera13a,Jacques2014,Acal20,Vidal2021}, canonical correlation \citep{Krzysko2013,Keser2015}, clustering \citep{Jacques2014b,Fortuna2018,Sharp21,Alva21} and  discriminant analysis \citep{Araki2009,Aguilera-Morillo20}, among other recent papers.  ANOVA problem for functional data (FANOVA) has also been considered in the literature. There are available different parametric \citep{Cuevas2004,CuestaFebrero2010} and non-parametric \citep{Delicado2007, Hall2007, jimenez2021a} approaches to tackle the traditional $m$-sample problem. A deep review of the most important aspects for FANOVA problem is developed in the book by \cite{Zhang2014}. An interesting comparison of multiple tests based on the idea of B-spline tests \citep{Shen04} can be seen in \cite{Gorecki15}. More recently, a novel approach based on functional principal component analysis was introduced in \cite{Aguilera2021}. The  purpose is to reduce the dimension of the problem and conduct a multivariate ANOVA on the vector of the most explicative principal component scores. Additionally, several authors have also focused their efforts on providing some tools to deal with the multivariate ANOVA problem for functional data \citep{Gorecki2017,Acal21}.

Despite its noticeable interest in  applications with real data, the repeated measures setting for functional data (FANOVA-RM)  is barely considered. This theoretical framework deals with the situation where a functional response variable is observed under different conditions (also called treatments) or time periods for each subject. The most of the works available in the literature only concerns the paired sample layout. The first statistic to solve this problem was introduced by \cite{Martinez2011}. This statistic is given by the integral of the difference between the sample mean functions, whose null distribution was approximated by considering multiple bootstrap and permutation methods.  Additionally, \cite{Smaga2019} developed a new way to approximate the distribution of the same statistic by means of Box-type approximation.  The performance of these approximations was very similar in relation to the sample size  and power, but the Box-type approximation proved to be faster from a computational viewpoint. However, this statistic only takes into account the between group variability. In order to solve the lack of information about the within group variability, two new statistics adapted from the classical paired $t$-test were introduced in \cite{Smaga2020}. These statistics were extended by considering the basis expansion of the sample curves  with the aim of detecting changes in air pollution during the COVID-19 pandemic in \cite{Acal21}. In the more general context of testing  homogeneity of paired  functional data,  a Cramér-von-Mises type test statistic is introduced in \cite{Ditzhaus2021} with application to  stock market returns.

The current work is focused on two-way FANOVA  problem, in which the subjects are classified in independent groups and the response variable is observed under different conditions for each individual. Thus, one factor represents the repeated measures effect (treatments) and the second one denotes the group contribution. \textcolor{red}{ This is the case of one of the aplications  developed in this paper (see Section 4.2) where the aim is to detect if there are significant differences in the kinetic curves (angle of knee)  recorded  under three different velocities (repeated measures) on  two independent samples (age groups). A simple FANOVA model can be expressed as follows in terms of the global mean function, main-effect functions and i.i.d. errors:
$$
x_{ijk}(t)=\mu(t)+\alpha_i(t)+\beta_j(t)+\epsilon_{ijk}(t)\ \forall t\in T,
$$
where $x_{ijk}(t)$ is the observed value of  the response variable for the $k$th subject in the $j$th age group,  measured under the $i$th running velocity   at moment $t$ in a continuous time interval $T,$ $(i=1,2,3; j=1,2; k=1,\dots ,n_j)$ with $n_j$ being the sample size for each age group.}

The main idea is to assume the basis expansion of the sample curves in order to turn the FANOVA-RM into a multivariate ANOVA with repeated measures, where the vector of basis coefficients of the sample curves would be the dependent multivariate variable. As far as we know, this theoretical setting has not been ever addressed in the literature, only in \cite{Martinez2011} it is briefly indicated  how the tests could be generalized for the case of more than two samples, but without further details about the case of independent groups. 

In addition to this introduction, the manuscript has four additional sections. The theoretical aspects related to the proposed methodology are in Section 2. An extensive Monte Carlo simulation study to show the good performance of two considered FANOVA-RM approaches is developed in Section 3. The applications with biomechanics data that motivate this study are presented in Section 4. Finally, the most important conclusions from this research are summarized in Section 5.

\section{Model set up}

As natural extension of the multivariate case, the aim of Functional Analysis of Variance is to estimate the effect of one or more factors on a functional response variable. In this paper, two factors are considered (two-way FANOVA model) which usually represent groups and treatment conditions. Two different basis expansion approaches are developed in what follows.

\subsection{Two-way FANOVA model}

Let $\lbrace x_{ijk}(t):  i=1,2,\ldots,m; j=1,2,\ldots,g; k=1,2,\ldots,n_{ij}; t\in T\rbrace$ denote $g\times m$  independent samples of curves defined on a continuous interval $T.$ That is, $x_{ijk}(t)$ represents the response variable of the $k$th subject in the $j$th group under $i$th treatment at instant $t.$ Note that each sample contains $n_{ij}$ observations and the total sample size is $n=\sum_{i=1}^m\sum_{j=1}^g n_{ij}.$  

Sample curves can be seen as observations of i.i.d. stochastic processes (functional variables) $\lbrace X_{ijk}(t): i=1,2,\ldots,m; j=1,2,\ldots,g;  k=1,2,\ldots,n_{ij};  t\in T \rbrace$ with distribution $SP(\mu_{ij} (t), \gamma(t,s)),$ with $\mu_{ij} (t)$ being the mean function and $\gamma (t,s)$ being the common covariance function associated with each of the $g\times m$  stochastic processes.  It is supposed that these stochastic processes are second order, continuous in quadratic mean and with sample functions in the Hilbert space $L^2[T]$ of squared integrable functions with the usual inner product 
\begin{equation*}
<f|g>=\int_T f(t)g(t)dt, \ \forall f,g\in L^2[T]. \label{Ex1}
\end{equation*}

In Two-way FANOVA model, functional data verify the functional linear model given by
\begin{equation}
	x_{ijk}(t)=\mu(t)+\alpha_i(t)+\beta_j(t)+\epsilon_{ijk}(t)\ \forall t\in T, \label{Ex2}
\end{equation}
where $\mu(t)$ is the overall mean function; $\alpha_i(t)$ and $\beta_j(t)$ are the $i$th and $j$th main-effect functions of treatments and groups, respectively; and $\epsilon_{ijk}(t)$ are i.i.d. errors with distribution $SP(0,\gamma(s,t))\ \forall  i=1,2,\ldots,m; j=1,2,\ldots,g; k=1,2,\ldots,n_{ij}.$ This model is the generalization of the model proposed by \cite{Zhang2014} when $n_{ij}=1.$ \\

In model (\ref{Ex2}), possible interactions between groups and treatments are assumed to be ignorable. This means that the effect of a certain factor's level is the same for each level of any other factor. In our case the effect of the treatments on the functional response would be the same for all groups.  The interaction functional parameter can be added in model (\ref{Ex2}) as follows
\begin{equation}
	x_{ijk}(t)=\mu(t)+\alpha_i(t)+\beta_j(t)+\theta_{ij}(t)+\epsilon_{ijk}(t), \ \forall t\in T.\label{Ex5}
\end{equation}

It is well known that FANOVA model is not identifiable (the functional parameters are not uniquely defined). This problem is overcome by applying certain constraints.  By extending the usual constraints in the balanced multivariate case (the cell sizes $n_{ij}$ are equal), the main effects and interaction effects functions sum to zero.  An appropriate sequence of positive weights must be considered to define the constraints in the unbalanced design \citep{Zhang2014}. From now on, we will assume the constraints   

\begin{eqnarray}
	&&\sum_{i=1}^m\alpha_{i}(t)=\sum_{j=1}^g\beta_{j}(t)=\sum_{i=1}^m \theta_{ij}(t)=\sum_{j=1}^g \theta_{ij}(t)=  \nonumber\\
	&&=\sum_{i=1}^m \sum_{j=1}^g \theta_{ij}(t)=0.
	\label{constraints}
\end{eqnarray}	

Under these constraints, we have that 
\begin{eqnarray}
	\alpha_{i}(t) & = & \mu_{i.} (t)- \mu (t) \nonumber \\ 
	\beta_{j}(t) & = & \mu_{.j} (t)- \mu (t) \nonumber\\
	\theta_{ij} (t) & = & \mu_{ij} (t) - \mu_{i.} (t) -\mu_{.j} (t) + \mu (t),
	\label{cons}
\end{eqnarray}
where $\mu_{i.} (t)$ and $\mu_{.j} (t)$ are the marginal mean functions of the functional response for each treatment and each group, respectively.

The most interesting hypothesis tests associated with two-way FANOVA model 
are given by the following null hypotheses against the alternative, in each case, that its negation holds:
\begin{itemize} 
	\item Testing if the main-effects of treatments are statistically significant (equality of the unknown treatments mean functions) 
	\begin{equation}
		H_0:\alpha_1(t)=\alpha_2(t)=\ldots=\alpha_m(t)=0, \ \forall t\in T, 
		\label{H1}
	\end{equation}
	\item Testing if the main-effects of groups are statistically significant (equality of the unknown groups mean functions)
	\begin{equation}
		 H_0:\beta_1(t)=\beta_2(t)=\ldots=\beta_g (t) = 0, \ \forall t\in T, 
		\label{H2}	
	\end{equation}
	\item Testing if the main-effects of the groups and treatments  are
	simultaneously null
	\begin{equation}
		H_0: \alpha_i (t) = \beta_j (t)= 0, \ \forall i,j; \ \forall t\in T,
		\label{H3}
	\end{equation}
	\item Testing if there is a significant interaction-effect between groups and treatments
	\begin{equation}
		H_0: \theta_{ij}(t)= 0, \ \forall i,j; \ \forall t\in T.
		\label{H4}
	\end{equation}
\end{itemize}

Different test statistics have been proposed in literature for FANOVA testing problems. A detailed study of point-wise F-test, L$^2$-norm-based test and functional F-type test under Gaussian assumption, together with $\chi^2$ and bootstrap approaches for non Gaussian samples can be seen in \cite{Zhang2014}.  Exhaustive simulation studies to compare multiple existing tests for one-way  FANOVA testing are presented in \cite{Gorecki15}. Different basis expansion and functional principal component approaches are proposed in \cite{Aguilera2021} and \cite{Aguilera2021Google} with applications in electronic and Google Trends, respectively. 

Under the constraints defined in (\ref{constraints}), the usual least squares estimators of the functional parameters in model (\ref{Ex5}) are obtained by minimizing
\begin{equation*}
\int_T \sum_{i=1} ^m \sum_{j=1}^g \sum_{k=1}^{n_{j}} [x_{ijk}(t) - (\mu(t)+\alpha_i (t)+\beta_j (t)+\theta_{ij}(t))]^2.
\end{equation*}

\subsection{Estimation and computation from basis expansions} 

In practice, functional data are not continuously observed over time but only discrete observations are available for each sample curve. What is more, the number of observations and the location of observed time points could be different for each curve. Because of this issue, the first step in FDA is to reconstruct the original functional form of the data by using some functional projection approach. 

Let us suppose that $\left \{ \phi_h (t) \right \}_{h=1, \dots, \infty }$ is a basis of the functional space $L^2[T]$ the curves belong to. Then, each curve admits an expansion into this basis as follows
\begin{equation}
	x_{ijk}(t)=\sum_{h=1}^\infty y_{ijkh}\phi_h(t), \label{Ex9}
\end{equation}
where the basis coefficients $y_{ijkh}$ are generated by random variables with finite variance. An approximated representation is usually obtained by truncating this basis expansion in terms of a number $p$ of basis functions sufficiently large to assure an accurate representation of each curve. 

From now on, it will be assumed that sample curves belong to the space generated by the basis $\left \{ \phi_1(t), \dots, \phi_p(t) \right \}.$ In vectorial form
\begin{equation*}
x_{ijk}(t)={\bf{y}}'_{ijk}{\bf{\Phi}}(t),
\end{equation*}
with ${\bf{y}}_{ijk}=(y_{ijk1},\ldots,y_{ijkp})'$ and ${\bf{\Phi}}(t)=(\phi_1(t),\ldots,\phi_p(t))'.$ This way, the initial curves $x_{ijk}$  are  replaced with their vectors of basis coefficients  ${\bf{y}}_{ijk}.$ The main advantage of this  approach is that the dimension of the data depends only on the number of curves and on the order $p$ of the expansion. Due to the fact that the curves are observed with error, some smoothing approach (e.g. least squares approximation) is usually performed to estimate the basis coefficients. In relation to the choice of a suitable basis, there are multiple options depending on the characteristics of the sample curves but the most common are Fourier, B-Splines or wavelets bases. Another key point is to select the dimension of the basis, which can be worked out, for instance, by generalized cross validation \citep{Craven79}. 

Hence, by assuming the basis expansion in (\ref{Ex9}), the estimators of the functional parameters can be expressed in terms of basis functions as follows
\begin{eqnarray}
	\hat{\mu}(t)& = & \overline{x}_{\cdot \cdot \cdot}(t)=\overline{{\bf{y}}}_{\ldots}'{\bf{\Phi}}(t) \nonumber \\
	\hat{\alpha}_{i}(t) & = & \overline{x}_{i \cdot \cdot}(t)-\overline{x}_{\cdot \cdot \cdot}(t) = (\overline{{\bf{y}}}_{i..}'-\overline{{\bf{y}}}_{\ldots}'){\bf{\Phi}}(t)
	\nonumber \\
	\hat{\beta}_{j}(t) & = & \overline{x}_{\cdot j \cdot}(t)-\overline{x}_{\cdot \cdot \cdot}(t) = (\overline{{\bf{y}}}_{.j.}'-\overline{{\bf{y}}}_{\ldots}'){\bf{\Phi}}(t)
	\nonumber \\
	\hat{\theta}_{ij}(t) & = & \overline{x}_{ij\cdot}(t)-\overline{x}_{i\cdot\cdot}(t)-\overline{x}_{\cdot j\cdot}(t)+\overline{x}_{\cdot \cdot \cdot}(t) = \nonumber\\
	&&(\overline{{\bf{y}}}'_{ij.}-\overline{{\bf{y}}}_{i..}'-\overline{{\bf{y}}}_{.j.}'+\overline{{\bf{y}}}_{\ldots}'){\bf{\Phi(t)}}  \nonumber \\
	\hat{\epsilon}_{ijk}(t) & = & x_{ijk}(t)-\overline{x}_{ij\cdot}(t) =
	({\bf{y}}_{ijk}'-\overline{{\bf{y}}}_{ij.}'){\bf{\Phi}}(t)
	\label{Ex6}
\end{eqnarray} 
where,
\begin{eqnarray}
	\overline{x}_{\cdot \cdot \cdot}(t) & = & {1\over m} \sum_{i=1}^m {1\over g} \sum_{j=1}^g {1\over n_{j}} \sum_{k=1}^{n_{j}} x_{ijk}(t) \nonumber \\
		\overline{x}_{ij\cdot}(t) & = & {1\over n_{j}} \sum_{k=1}^{n_{j}} x_{ijk}(t), \ \ i=1,\ldots,m; \ j=1,\ldots,g \nonumber \\
	\overline{x}_{i \cdot \cdot}(t) & = & {1\over g} \sum_{j=1}^g {1\over n_{j}} \sum_{k=1}^{n_{j}} x_{ijk}(t), \ \ i=1,\ldots,m  \nonumber \\
	\overline{x}_{\cdot j \cdot}(t) & = &{1\over m} \sum_{i=1}^m {1\over n_{j}} \sum_{k=1}^{n_{j}}x_{ijk}(t), \ \ j=1,\ldots,g 
	\label{Ex7}
\end{eqnarray} 
with $\overline{{\bf{y}}}_{\ldots}$, $\overline{{\bf{y}}}_{i..}$, $\overline{{\bf{y}}}_{.j.}$ and $\overline{{\bf{y}}}_{ij.}$ being the grand mean vector, the treatment mean vector, the group mean vector and the interaction mean vector, respectively, associated with the vector of sample curves basis coefficients ${\bf{y}}_{ijk}$. 

The results above proves that Two-Way FANOVA model turns into Two-Way multivariate ANOVA model for the $p$-dimensional response variable   $(Y_1, Y_2, \dots, Y_p)$ that generates the basis coefficients of the functional variable $X.$ 

\subsection{Repeated measures approaches}

Two-Way FANOVA model presented above corresponds with independent samples. Nevertheless,  in many fields such as medicine, social sciences, education or psychology, among others, it is very common to deal with a repeated measures design in  which measurements on one or more response variables are conducted at several occasions (longitudinal data) or under different treatment conditions on the same subject. When a single response variable is observed, the design is called univariate repeated measures design. In this document, this approach is extended to test the effect of two factors (treatment and group) on a functional random variable. This new approach is known as Two-Way FANOVA-RM. Obviously, the simplest approach would be the model where the treatment factor is only considered. This model can be derived through the model with two factors. 

Let us now suppose that we have a repeated measures design with $g$ independent samples of curves (one per group) so that the response functional variable $X$ is repeatedly measured on each subject at $m$ different time periods (longitudinal functional data) or under $m$ different treatment conditions.

Let $\lbrace x_{ijk}(t):  i=1,2,\ldots,m; j=1,2,\ldots,g; k=1,2,\ldots,n_{j}; t\in T\rbrace$ denote $g$ independent samples of curves defined on a continuous interval $T.$ That is, $x_{ijk} (t)$ is the response of the $k$th subject in the $j$th group under  the $i$th treatment. Note that the response of each subject is observed $m$ times so that we have $n=\sum_{j=1}^g n_j$ subjects and $n\times m$ sample curves. It is assumed that each treatment is applied to all subjects (balanced design). This fact will be an essential aspect later.

The main objective of this manuscript is to adapt Two-Way FANOVA model to the case of repeated measures by taking the intra-subject variability into account. We propose to perform a multivariate analysis of variance with repeated measures on the multivariate response defined by the random basis coefficients of the functional variable.

As far as the authors know, there are two different models to include the intra-subject effect in the analysis: Doubly Multivariate Model (DMM) and Mixed Multivariate Model (MMM). Both assume the multivariate normality hypothesis and homogeneity of covariance matrices. The difference between them arises in the assumptions on the covariance matrix. DMM only assumes that the covariance matrix is positive definite, whereas MMM requires the multivariate sphericity condition. This restrictive condition is not verified in many real situations, so that DMM is more frequently used. However, if the sphericity condition is satisfied, MMM should be employed because it is more powerful \citep{Bock1975}. Several reviews, new results and comparisons of both models by standing out their principal characteristics and behaviour on applications were developed in \cite{Timm1980} and \cite{Boik1988,Boik1991}. An application with data from an educational survey can be seen in \cite{Filiz2003}. Three different methods to solve the lack of variance homogeneity are studied in \cite{LixLloyd2007}. Finally, a new statistic based on DMM is developed in \cite{Hirunkasi2011} for the tricky scenario where the dimension of the response variable is greater than the number of observations.

\subsubsection{Doubly Multivariate Model}
In our functional data context, FANOVA-RM model can be written as a MANOVA-RM model for the basis coefficients of the sample curves as
\begin{equation}
	\bf{Y}=\bf{\mathcal{X}B}+\bf{E}, \label{Ex12}
\end{equation}
\noindent where the response $\bf{Y}$ is the matrix $n\times (p\times m)$ whose rows contains the $p$-dimensional  basis coefficients of the functional response variable $X$ for  the $n$ subjects (distributed amongst $g$ independent samples) examined under each of the $m$ treatments ($p\times m$ dimensional response ordered within each row according to treatment an within treatment according to the basis coefficients). Les us observe that in this model we have $p\times m$ response variables that represent the $p$-dimensional vector of basis coefficients for each treatment. 
In addition, $\bf{\mathcal{X}}$ is the between group design matrix $n\times g$ and  $\bf{B}$ is the unknown parameter matrix $g\times (p\times m)$, with $\bf{E}$ being the error matrix $n\times (p\times m)$ whose  rows $E_i$ are i.i.d. $N_{pm} (\bf{0}, \bf{\Sigma}),$ so that
\begin{equation}
	vec(E')\sim N_{npm}(\bf{0},   \bf{I}_n \otimes \bf{\Sigma} ),
	\label{normal}
\end{equation} 
where $\bf{I}_n$ is the identity matrix and $\bf{\Sigma}$ is a $(p\times m)\times (p\times m)$ positive definite matrix. 

The hypotheses tests of interest related with the statistical significance of treatment, group and interaction effects, i.e. (\ref{H1}), (\ref{H2}) and (\ref{H4}), respectively, can be expressed in terms of basis functions and formulated through the following general linear hypothesis
\begin{equation}
	H_0: \bf{G}' \bf{B}(\bf{T} \otimes \bf{I}_p)={\bf{0}}, \label{Ex13}
\end{equation}
where ${\bf{0}}$ is a matrix of zeros with appropriate order, $\bf{G}$ is a matrix $g\times s$ (rank $s$) which contains  the coefficients for between group tests and $\bf{T}$ is a matrix $m\times q$ (rank $q$) which contains the coefficients  for within treatments tests. The columns of the matrix $\bf{G}$ are composed by the coefficients of $s$ estimable between group functions and the columns of the matrix $\bf{T}$ are the coefficients of $q$ linear functions of the $m$ treatments. Without loss of generality, it is assumed that $\bf{T}$ is chosen to be orthonormal $\bf{T}'\bf{T}=\bf{I}.$  Depending on the type of contrast and the objective of the study, matrices $\bf{G}$ and $\bf{T}$ will have different expressions. Deep studies related with these topics were developed in  \cite{Timm1980}, \cite{Thomas1983},   \cite{HandTaylor1987} and \cite{Timm2002}.

A DMM testing problem is worked out by means of the usual MANOVA statistics, e.g., Wilks's lambda (W), Lawley-Hotelling's trace (LH), Pillai's trace (P) or Roy's maximum root (R), associated with the following reduced $qp$-dimensional multivariate linear model 
\begin{equation}
	\bf{Y}(\bf{T} \otimes \bf{I}_p)=\bf{\mathcal{X}B}(\bf{T} \otimes \bf{I}_p)+\bf{E}(\bf{T} \otimes \bf{I}_p). \label{Ex14}
\end{equation}

These MANOVA statistics are based on the relation between the sum of squares and cross product matrices corresponding to error and hypothesis obtained by
\begin{eqnarray*}
	\bf{S}_e&=&(\bf{T}' \otimes \bf{I}_p)\bf{Y}'[\bf{I}_n-\bf{\mathcal{X}}(\bf{\mathcal{X}}'\bf{\mathcal{X}})^{-}\bf{\mathcal{X}}']\bf{Y}(\bf{T} \otimes \bf{I}_p)\\
	 \bf{S}_h&=&(\bf{T}' \otimes \bf{I}_p)  \bf{\hat{B}}' \bf{G} [\bf{G}' (\bf{\mathcal{X}}'\bf{\mathcal{X}})^{-} \bf{G}]^{-1}\bf{G}'\bf{\hat{B}}(\bf{T} \otimes \bf{I}_p),\label{Ex15}
\end{eqnarray*} 
where $\bf{\hat{B}}$ is the maximum likelihood estimator of $\bf{B}$ given by $\bf{\hat{B}} = (\bf{\mathcal{X}}' \bf{\mathcal{X}})^{-} \bf{\mathcal{X}}' \bf{Y},$ with $(\bf{\mathcal{X}}'\bf{\mathcal{X}})^{-}$ being any generalized inverse of $\bf{\mathcal{X}}'\bf{\mathcal{X}}.$  In practice, W, LH, P and R are usually approximated by F-tests statistics through Rao's approximation \citep{Rencher12}. Finally, it is important to keep in mind that DMM can only be used when $n>p\times m$, since otherwise the matrix $\bf{S}_e$ would be singular.

\subsubsection{Mixed Multivariate Model} 

A functional mixed-effect model can be considered in order to model the intra-subject variability. The expression of Two-Way mixed-effect FANOVA model is similar to (\ref{Ex5}), except for  the inclusion of a new random subject-effect functional parameter, $\pi_k(t)$, in the model, $\forall t\in T,$ as 
\begin{equation*}
	x_{ijk}(t)=\mu(t)+\alpha_i(t)+\beta_j(t)+\theta_{ij}(t)+\pi_k (t)+\epsilon_{ijk}(t), \label{Ex11}
\end{equation*}
where  $\pi_k(t)$ are i.i.d.  subject-effects with distribution $SP(0,\gamma_\pi(s,t)),$ and  $\epsilon_{ijk}(t)$ are  i.i.d. errors with distribution $SP(0,\gamma_{\epsilon}(s,t))\ \forall  i=1,2,\ldots,m; j=1,2,\ldots,g; k=1,2,\ldots,n_{j}.$ Besides, $\pi_k(t)$ and $\epsilon_{ijk}(t)$ are mutually independent so that $\gamma (t,s) = \gamma_\pi(s,t)+\gamma_\epsilon(s,t).$

The multivariate  mixed-effect model is a generalization of Scheffé's Univariate Mixed Model \citep{Scheffe1956}. In our FANOVA-RM approach, the model for the $p$-dimensional response of basis coefficients can be expressed as 
\begin{equation*}
{\boldsymbol{y}}_{ijk}={\boldsymbol{\mu}}+{\boldsymbol{\alpha}}_i+{\boldsymbol{\beta}}_j+\boldsymbol{\theta}_{ij}+{\boldsymbol{\pi}}_k+{\boldsymbol{\epsilon}}_{ijk}, \label{Ex16}
\end{equation*}
where ${\boldsymbol{\pi}}_k$ are i.i.d. subject effects with distribution $N(0,\bf{\Sigma}_{\pi})$ and ${\boldsymbol{\epsilon}}_{ijk}$ are i.i.d. errors with distribution $N(0,\bf{\Sigma}_{\epsilon}).$ In addition, the covariance matrix of the $p$-dimensional response is given by $\bf{\Sigma} =\bf{\Sigma}_{\pi} +\bf{\Sigma}_{\epsilon}$ because $\pi_k(t)$ and $\epsilon_{ijk}(t)$ are mutually independent. MMM  can be expressed  in terms of the linear model for multivariate repeated measures defined in (\ref{Ex12}) by rearranging  the data matrix $\bf{Y}$ as follows.

Let us denote by $\bf{y}_i$ the  {\it ith} row of the response matrix $\bf{Y}$ in model (\ref{Ex12}) that represents the $pm$-dimensional vector of responses values for the $i$th sample subject.  Then, the vector $\bf{y}_i$ is rearranged to obtain a $m\times p$ matrix  $\bf{Y}_i^*$ such that $vec((\bf{Y}_i^*)') = \bf{y}_i.$ The $vec()$-operator stacks the columns of a matrix. Thus, the rows and columns of $\bf{Y}_i^*$ correspond with the treatments and the dependent variables, respectively. Considering this transformation, the rearranged response matrix for MMM analysis is
\begin{equation*}
	\bf{Y}^*= \left (
	\begin{array}{l}
		\bf{Y}_1^* \\
		
		\dots \\
		\bf{Y}_n^* \\
	\end{array}
	\right ).
	\label{Ex17}
\end{equation*}

If $\bf{B}$ and $\bf{E}$ are rearranged in the same way, a $(g\times m)\times p$  matrix of unknown parameters and a $(n\times m)\times p$ matrix of random errors are obtained. These rearranged matrices verify that $vec(\bf{Y}^{*'}) = vec(\bf{Y}'),
vec(\bf{B}^{*'}) = vec(\bf{B}'), vec(\bf{E}^{*'}) = vec(\bf{E}').$

After this transformation, (\ref{Ex12}) and (\ref{Ex14}) can be written as
\begin{equation*}
\bf{Y}^*=(\bf{\mathcal{X}}\otimes \bf{I}_m) \bf{B}^* +\bf{E}^* \label{Ex18}
\end{equation*}
\begin{equation}
	(\bf{I}_n \otimes \bf{T}') \bf{Y}^*=  (\bf{\mathcal{X}}\otimes \bf{T}') \bf{B}^* + (\bf{I}_n \otimes \bf{T}') \bf{E}^*.
	\label{Ex19}
\end{equation}

Let us observe that the rows of the error matrix $\bf{E}^*$ are normally distributed but not independent because the rows corresponding to the observation of the response variable on the same  individual under the different treatments conditions could be correlated. Therefore, (\ref{Ex19}) is a mixed model that keep the influence of the individuals on the response variable in mind (individual random effects).

Now, the general null hypothesis of interest is given by
\begin{equation}
	H_0:  (\bf{G}' \otimes \bf{T}') \bf{B}^* ={\bf{0}}, \label{nullMMM}
\end{equation}
that is the same than (\ref{Ex13}). Then,  the matrices corresponding to error and hypothesis for testing (\ref{nullMMM}) are 
\begin{eqnarray*}
	 \bf{S}_e^*&=&\bf{Y}^{*'}[(\bf{I}_n-\bf{\mathcal{X}}(\bf{\mathcal{X}}'\bf{\mathcal{X}})^{-}\bf{\mathcal{X}}')\otimes \bf{TT}']\bf{Y}^{*}\\
	 \bf{S}_h^*&=&\bf{Y}^{*'}\bf{\Upsilon} \bf{Y}^{*},\label{Ex20}
\end{eqnarray*} 
with $\bf{\Upsilon}=(\bf{\mathcal{X}}(\bf{\mathcal{X}}'\bf{\mathcal{X}})^{-}\bf{G}'[\bf{G}(\bf{\mathcal{X}}'\bf{\mathcal{X}})^{-}\bf{G}']^{-1}\bf{G}(\bf{\mathcal{X}}'\bf{\mathcal{X}})^{-}\bf{\mathcal{X}}')\otimes \bf{TT}'.$

For a MMM to be valid it must be verified that the $\bf{S_e}$ and $\bf{S_h}$ matrices have to be independently distributed  as Wishart matrices. Independence is derived from multivariate normality and homogeneity of the errors given in (\ref{normal}) but a new assumption on the covariance structure, called multivariate sphericity, is a necessary and suffcient condition for these matrices to be Wishart.  Note that sphericity is a situation more general of the composed symmetry. A likelihood ratio test for checking the multivariate sphericity is derived in \cite{Thomas1983} but the asymptotic distribution may be a poor approximation when the sample size is moderate. An approximation that solves the lack of power for moderate sample sizes is developed in \cite{Boik1988} by applying Box's expansion of the characteristic function \citep{Box1949}. For the univariate case, \cite{Box1954} proposed a factor of correction with the goal of giving a solution when the sphericity is not verified. This method consisted of disrupting the degrees of freedom  of F-statistic. In this line, \cite{Boik1988} formulated an analogous approach  for the multivariate case.

From a theoretical viewpoint, by considering the restricted data matrix $\bf{Y}(\bf{T} \otimes \bf{I}_p)$ that examines $q$ functions of the treatments, multivariate sphericity is a condition for the structure of the covariance matrix $\bf{\Omega}$ of $\bf{Y}(\bf{T} \otimes \bf{I}_p)$. Thus, $\bf{\Omega} = Cov(\bf{Y}(\bf{T} \otimes \bf{I}_p)) = (\bf{T}' \otimes \bf{I}_p) \bf{\Sigma} (\bf{T} \otimes \bf{I}_p).$ In particular, this condition assumes that $\bf{\Omega} = \bf{I}_q \otimes \bf{\Gamma}$ with  $\bf{\Gamma}$ being a $p \times p$ positive definite matrix of covariances among the $p$ response variables. The variation of the combinations of treatment levels is captured in the $(q\times p) \times p$ diagonal blocks of $\bf{\Omega}.$ In this sense, multivariate sphericity gives raise that all these blocks are identical and that the $q$ subvectors of each row of restricted data matrix are independent. Thus, multivariate sphericity can be seen as a condition about the variation of the dissimilarities between treatment modalities.

Finally, let us observe that MMM only can be used when $n\times m>p$, although this assumption is almost always fulfilled in practice.

\section{Simulation study}

In this section, an extensive Monte Carlo simulation study is carried out in order to test the performance of Two-Way FANOVA-RM methods proposed in previous section. Specifically, four different simulation studies are developed with the purpose of evaluating the behaviour of the tests by considering different type of errors, sample sizes and shapes of the functions that provide the curves. Sample curves are generated artificially according to the following functional mixed ANOVA model
\begin{eqnarray*}
x_{ijk}(t)&=&\alpha_i(t)+\beta_j(t)+\theta_{ij}(t)+\gamma_k sin(\pi t)+\epsilon_{ijk}(t),  \\
&&i=1,2,3; j=1,2; k=1,...,n_j,
\end{eqnarray*}
where $\alpha_i(t)$ and $\beta_j(t)$ are the $i$th and $j$th main-effect functions of treatments and groups, respectively; $\theta_{ij}(t)$ is the $(i,j)$th interaction-effect between treatments and groups; $\gamma_ksin(\pi t)$ represents the subject-effect and $\epsilon_{ijk}(t)$ is the error function.

In the four studies, functional data have been generated in discretized versions $x_{ijk}(t_r)$ for $r=1,...,101$ with $t_1,...,t_{101}$ being chosen equidistant in the interval [0,1]. Least squares approximation in terms of a basis of cubic B-splines with 14 functions was employed in all cases in order to reconstruct the functional form of sample curves. Therefore, the sample of each of the three treatments ($m=3$) is represented by a vector of  14 dependent variables ($p=14$). Besides, it is considered $n_1=n_2=n$ with $n=50,100$ to check the power of the tests when \def\stacktype{S} 
$n> (\approx)\ p\times m$ and $n\gg p\times m$. Inspired by \cite{durban2005}, the random subject-effect in all models is given by $\gamma_k \sim N(\mu_k,\sigma_k=0.2)$ with $\mu_k\sim U(0,0.05).$

The tests were replicated 500 times for each of the scenarios to be specified below. Significance level was established as $\alpha=0.05$. Finally, Wilks' Lambda statistics was conducted both DMM and MMM for testing if the profiles for each variable are parallel and whether there are differences in treatments and in groups. For $\bf{G}$ and $\bf{T}$ from (\ref{Ex13}), the following matrices were employed
$$
\begin{array}{l}
	\ \bf{G}=\begin{pmatrix}
		1 & -1
	\end{pmatrix}, \ \bf{T}=\begin{pmatrix}
		\frac{1}{\sqrt{2}} & \frac{-1}{\sqrt{6}} \\ 
		0 & \frac{2}{\sqrt{6}}\\ 
		\frac{-1}{\sqrt{2}} & \frac{-1}{\sqrt{6}}
	\end{pmatrix} \ \mathrm{for \ interaction \ test;}\\
	\ \\
	\ \bf{G}=\begin{pmatrix}
		1 & -1
	\end{pmatrix}, \ \bf{T}=\bf{I}_3 \ \mathrm{for \ group \ test;}\\
	\ \\
	\ \bf{G}=\bf{I}_2 , \ \bf{T}=\begin{pmatrix}
		\frac{1}{\sqrt{2}} & \frac{-1}{\sqrt{6}} \\ 
		0 & \frac{2}{\sqrt{6}}\\ 
		\frac{-1}{\sqrt{2}} & \frac{-1}{\sqrt{6}}
	\end{pmatrix} \ \mathrm{for \ treatment \ test}.\\
\end{array}
$$

\subsection{First scenario (M1)}
Three different forms are assumed for the main-effect functions of treatments and groups, and two for the interaction-effect functions. Thus, 18 different models are obtained by making all possible combinations among them. The selection of these functions was inspired by \cite{Cuevas2004}, \cite{Gorecki15} and \cite{Gorecki2017} and they are given by

\begin{enumerate}
	\item[] M1.A1: $\alpha_i(t)=t(1-t), \ i=1,2,3;$	
	\item[] M1.A2: $\alpha_i(t)=t^{i/5}(1-t)^{6-i/5}, \ i=1,2,3;$	
	\item[] M1.A3: $\alpha_i(t)=t^i(1-t)^{6-i},\  i=1,2,3;$
	\item[] M1.B1: $\beta_j(t)=0.1\times |sin(4\pi t)|, j=1,2;$		
	\item[] M1.B2: $\beta_j(t)=(0.05\times j)\times |sin(4\pi t)|,\ j=1,2;$
	\item[] M1.B3: $\beta_j(t)=(0.025\times j)\times|sin(4\pi t)|,\  j=1,2;$
	\item[] M1.I1: $\theta_{ij}(t)=[sin(2\pi t^2)]^5, \  i=1,2,3, j=1,2;$		
	\item[] M1.I2: $\theta_{ij}(t)=[sin(2\pi t^2)]^{5+2ij},\   i=1,2,3, j=1,2.$
\end{enumerate}

The null hypothesis of interest holds for models with M1.A1, M1.B1 and M1.I1, whereas the opposite happens for the remainder. In order to evaluate the results, it is important to keep in mind that the interaction function M1.I2 changes for each level of treatment and group.  The main differences between M1.A2 and M1.A3 (the same for M1.B2 and M1.B3) are that the main effects in M1.A2 and M1.B2 are quite separated so that the testing problem should be less harder. Finally, $\epsilon_{ijk}(t_r)$ are i.i.d. random variables $N(0,\sigma_\epsilon)$ with $\sigma_{\epsilon}=0.10,0.20,0.40$. The latter value, i.e. $\sigma_\epsilon=0.40$, is introduced for checking the performance of the tests under extreme situations. \textcolor{red}{Figure \ref{curvesSIMU} displays the differences among main effects  functions  when the treatment, group and interaction functions are considered to be different}. Table \ref{tablaM1} shows the acceptance proportions for each scenario. The obtained outcomes can be summed up in the following commentaries:

\begin{figure*}
\hspace*{-0.25cm}
    {\includegraphics[width=6.1cm, height=6.1cm]{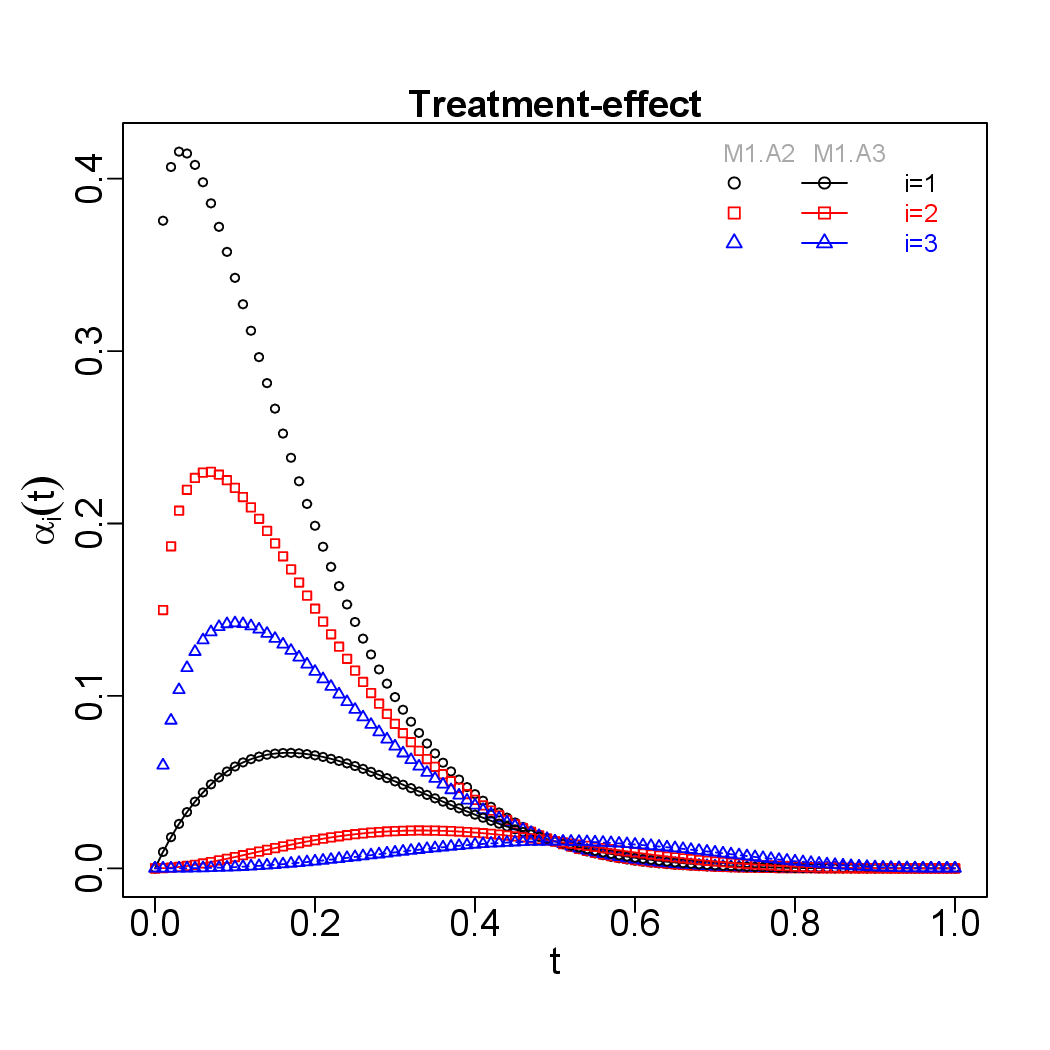}}\hspace*{-0.3cm}
    {\includegraphics[width=6.1cm, height=6.1cm]{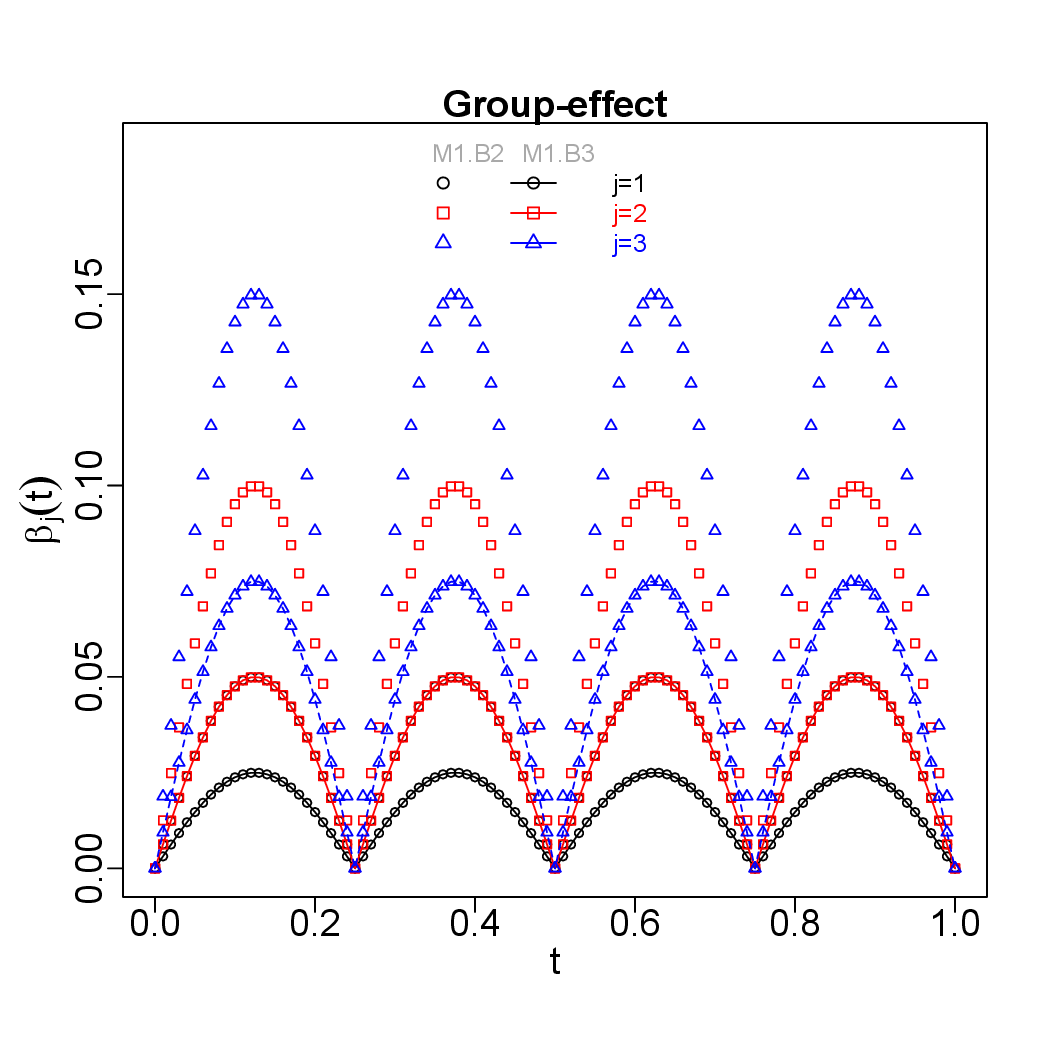}}\hspace*{-0.3cm}
    {\includegraphics[width=6.1cm, height=6.1cm]{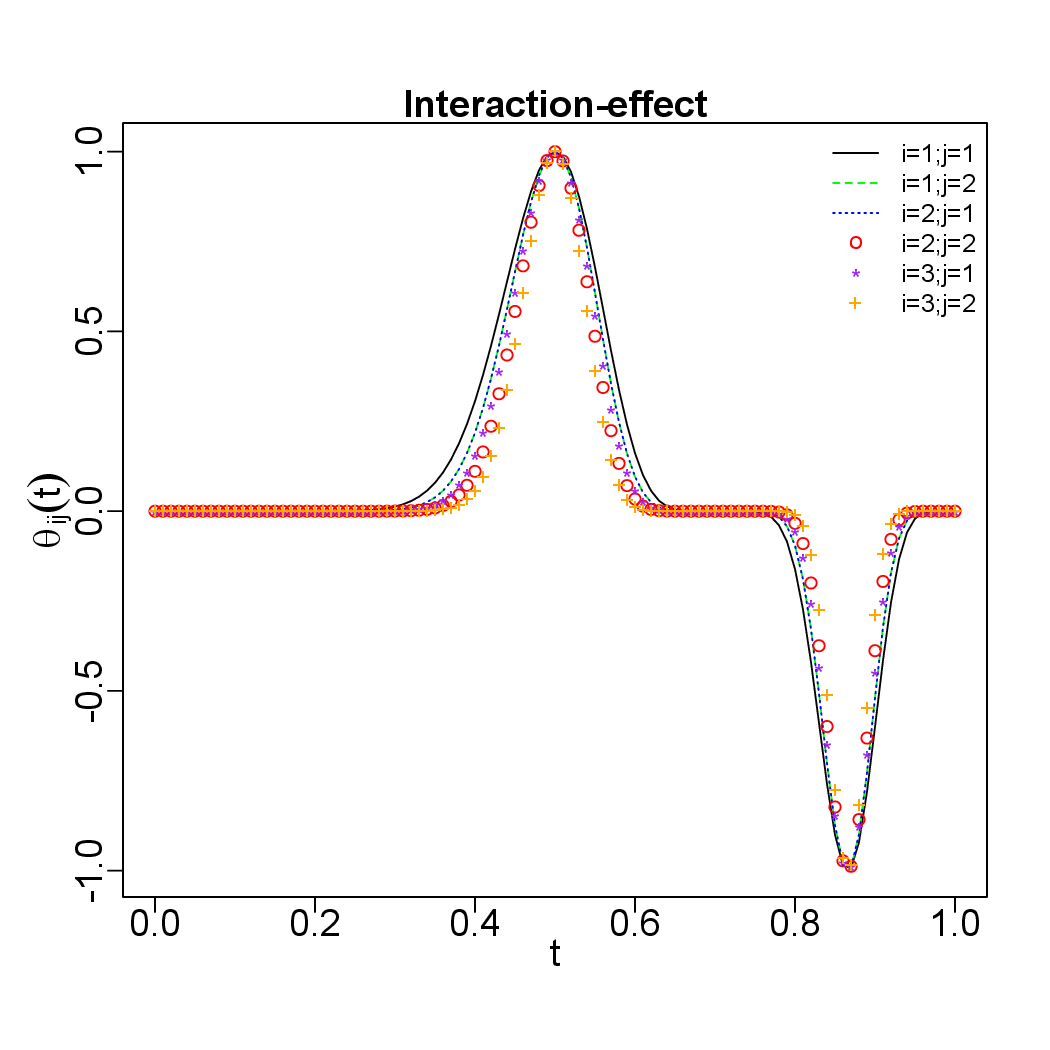}}
	\caption{\textcolor{red}{Main-effect  (treatments and groups) and  interaction-effect functions (first and second scenarios).}}
	\label{curvesSIMU}
\end{figure*}

\begin{enumerate}
	\item For the cases where the three null hypothesis are true, the tests reach good results (the error rates are lower than 7\% in all cases). Furthermore, for the rest of cases,  the decision of the test when $\sigma_\epsilon=0.10$ is really satisfactory even in the more extreme situations.
	\item MMM gets better results than  DMM, being the differences in some occasions almost of 20\%.  In this context, as long as the MMM's conditions are satisfied is better to use this approach.
	\item The sample size and the dispersion parameter play an important role in the analysis. As the value of $\sigma_\epsilon$ increases the error rate turn into larger, especially when the sample size is similar to the dimension of response variables.
	\item Regarding the check of differences among treatments, the tests provide adequate outcomes. It is only found out an error rate a little higher when jointly $n=50$ and $\sigma_\epsilon=0.40$  for M1.A3 (the acceptance proportions vary between 0.340-0.392 and 0.244-0.260 in DMM  and MMM, respectively). Remind that in M1.A3 the differences between the treatments functions are smaller. Secondly, in relation to the differences among groups, the performance of the tests is suitable except for $\sigma_\epsilon=0.40$. In this case  good results are appreciated only when $n=100$ and the similarities between the groups are quite separated (M1.B2). It is also discovered a great error rate when $n=50$, $\sigma_\epsilon=0.20$ and M1.B3 were considered together. Finally, the power of the tests for the difference of the interactions between groups and treatment is too small when, on the one hand, $\sigma_\epsilon=0.40$, and on the other hand, $\sigma_\epsilon=0.20$ and $n=50$ are considered at the same time. It is worrisome the frequency of a correct decision in these situations.
\end{enumerate}

	\begin{sidewaystable*}
	\centering
	\caption{Observed acceptance proportions for each scenario at a significance level 0.05 in the case of simulation study M1.}\label{tablaM1}
	\resizebox{23cm}{!} {
		\begin{tabular}{|c|c|c|c|ccc|ccc|ccc|ccc|ccc|ccc|} 
			\cline{5-22}
			\multicolumn{1}{c}{}              & \multicolumn{1}{c}{}             & \multicolumn{1}{c}{} &          & \multicolumn{6}{c|}{$\sigma_\epsilon$=0.10}                & \multicolumn{6}{c|}{$\sigma_\epsilon$=0.20}                & \multicolumn{6}{c|}{$\sigma_\epsilon$=0.40}                 \\ 
			\cline{5-22}
			\multicolumn{1}{c}{}              & \multicolumn{1}{c}{}             & \multicolumn{1}{c}{} &          & \multicolumn{3}{c|}{M1.I1} & \multicolumn{3}{c|}{M1.I2} & \multicolumn{3}{c|}{M1.I1} & \multicolumn{3}{c|}{M1.I2} & \multicolumn{3}{c|}{M1.I1} & \multicolumn{3}{c|}{M1.I2}  \\ 
			\cline{3-22}
			\multicolumn{1}{c}{}              &                                  & n                    & Approach & Parall. & Treat. & Group   & Parall. & Treat. & Group   & Parall. & Treat. & Group   & Parall. & Treat. & Group   & Parall. & Treat. & Group   & Parall. & Treat. & Group    \\ 
			\hline
			\multirow{12}{*}{\rotcell{M1.A1}} & \multirow{6}{*}{\rotcell{M1.B1}} & \multirow{2}{*}{50}  & Doubly   & 0.942   & 0.946  & 0.944   & 0       & 0      & 0       & 0.956   & 0.966  & 0.954   & 0.504   & 0      & 0       & 0.950   & 0.954  & 0.948   & 0.868   & 0      & 0        \\
			&                                  &                      & Mixed    & 0.956   & 0.946  & 0.956   & 0       & 0      & 0       & 0.954   & 0.962  & 0.944   & 0.404   & 0      & 0       & 0.946   & 0.948  & 0.930   & 0.866   & 0      & 0        \\
			&                                  & \multirow{2}{*}{100} & Doubly   & 0.964   & 0.964  & 0.944   & 0       & 0      & 0       & 0.940   & 0.956  & 0.946   & 0.064   & 0      & 0       & 0.944   & 0.952  & 0.952   & 0.728   & 0      & 0        \\
			&                                  &                      & Mixed    & 0.950   & 0.970  & 0.952   & 0       & 0      & 0       & 0.970   & 0.950  & 0.952   & 0.054   & 0      & 0       & 0.948   & 0.960  & 0.956   & 0.702   & 0      & 0        \\ 
			\cline{2-22}
			& \multirow{6}{*}{\rotcell{M1.B2}} & \multirow{2}{*}{50}  & Doubly   & 0.956   & 0.952  & 0       & 0.002   & 0      & 0       & 0.954   & 0.948  & 0.014   & 0.528   & 0      & 0       & 0.948   & 0.946  & 0.612   & 0.862   & 0      & 0        \\
			&                                  &                      & Mixed    & 0.948   & 0.966  & 0       & 0       & 0      & 0       & 0.956   & 0.954  & 0       & 0.414   & 0      & 0       & 0.936   & 0.956  & 0.378   & 0.870   & 0      & 0        \\
			&                                  & \multirow{2}{*}{100} & Doubly   & 0.950   & 0.950  & 0       & 0       & 0      & 0       & 0.956   & 0.960  & 0       & 0.092   & 0      & 0       & 0.958   & 0.968  & 0.150   & 0.750   & 0      & 0        \\
			&                                  &                      & Mixed    & 0.960   & 0.970  & 0       & 0       & 0      & 0       & 0.942   & 0.946  & 0       & 0.036   & 0      & 0       & 0.942   & 0.954  & 0.052   & 0.736   & 0      & 0        \\ 
			\cline{2-22}
			& \multirow{6}{*}{\rotcell{M1.B3}} & \multirow{2}{*}{50}  & Doubly   & 0.950   & 0.956  & 0       & 0       & 0      & 0       & 0.948   & 0.946  & 0.612   & 0.532   & 0      & 0       & 0.948   & 0.944  & 0.908   & 0.872   & 0      & 0        \\
			&                                  &                      & Mixed    & 0.950   & 0.968  & 0.008   & 0       & 0      & 0       & 0.954   & 0.956  & 0.466   & 0.394   & 0      & 0       & 0.946   & 0.946  & 0.890   & 0.854   & 0      & 0        \\
			&                                  & \multirow{2}{*}{100} & Doubly   & 0.952   & 0.960  & 0       & 0       & 0      & 0       & 0.952   & 0.942  & 0.146   & 0.086   & 0      & 0       & 0.960   & 0.958  & 0.778   & 0.740   & 0      & 0        \\
			&                                  &                      & Mixed    & 0.958   & 0.956  & 0       & 0       & 0      & 0       & 0.942   & 0.936  & 0.088   & 0.062   & 0      & 0       & 0.958   & 0.948  & 0.714   & 0.678   & 0      & 0        \\ 
			\hline
			\multirow{12}{*}{\rotcell{M1.A2}} & \multirow{6}{*}{\rotcell{M1.B1}} & \multirow{2}{*}{50}  & Doubly   & 0.950   & 0      & 0.954   & 0       & 0      & 0       & 0.936   & 0      & 0.930   & 0.558   & 0      & 0       & 0.946   & 0      & 0.952   & 0.882   & 0      & 0        \\
			&                                  &                      & Mixed    & 0.946   & 0      & 0.934   & 0       & 0      & 0       & 0.960   & 0      & 0.944   & 0.414   & 0      & 0       & 0.952   & 0      & 0.930   & 0.890   & 0      & 0        \\
			&                                  & \multirow{2}{*}{100} & Doubly   & 0.952   & 0      & 0.952   & 0       & 0      & 0       & 0.958   & 0      & 0.947   & 0.098   & 0      & 0       & 0.950   & 0      & 0.950   & 0.736   & 0      & 0        \\
			&                                  &                      & Mixed    & 0.948   & 0      & 0.942   & 0       & 0      & 0       & 0.970   & 0      & 0.972   & 0.054   & 0      & 0       & 0.962   & 0      & 0.946   & 0.720   & 0      & 0        \\ 
			\cline{2-22}
			& \multirow{6}{*}{\rotcell{M1.B2}} & \multirow{2}{*}{50}  & Doubly   & 0.942   & 0      & 0       & 0.002   & 0      & 0       & 0.940   & 0      & 0.012   & 0.496   & 0      & 0       & 0.946   & 0      & 0.638   & 0.866   & 0      & 0        \\
			&                                  &                      & Mixed    & 0.948   & 0      & 0       & 0.002   & 0      & 0       & 0.950   & 0      & 0       & 0.464   & 0      & 0       & 0.950   & 0      & 0.420   & 0.840   & 0      & 0        \\
			&                                  & \multirow{2}{*}{100} & Doubly   & 0.950   & 0      & 0       & 0       & 0      & 0       & 0.962   & 0      & 0       & 0.082   & 0      & 0       & 0.956   & 0      & 0.140   & 0.734   & 0      & 0        \\
			&                                  &                      & Mixed    & 0.960   & 0      & 0       & 0       & 0      & 0       & 0.954   & 0      & 0       & 0.060   & 0      & 0       & 0.946   & 0      & 0.074   & 0.724   & 0      & 0        \\ 
			\cline{2-22}
			& \multirow{6}{*}{\rotcell{M1.B3}} & \multirow{2}{*}{50}  & Doubly   & 0.948   & 0      & 0.016   & 0       & 0      & 0       & 0.940   & 0      & 0.630   & 0.498   & 0      & 0       & 0.936   & 0      & 0.868   & 0.878   & 0      & 0        \\
			&                                  &                      & Mixed    & 0.940   & 0      & 0       & 0       & 0      & 0       & 0.946   & 0      & 0.456   & 0.400   & 0      & 0       & 0.932   & 0      & 0.794   & 0.854   & 0      & 0        \\
			&                                  & \multirow{2}{*}{100} & Doubly   & 0.938   & 0      & 0       & 0       & 0      & 0       & 0.938   & 0      & 0.138   & 0.084   & 0      & 0       & 0.934   & 0      & 0.766   & 0.698   & 0      & 0        \\
			&                                  &                      & Mixed    & 0.956   & 0      & 0       & 0       & 0      & 0       & 0.932   & 0      & 0.082   & 0.068   & 0      & 0       & 0.952   & 0      & 0.686   & 0.710   & 0      & 0        \\ 
			\hline
			\multirow{12}{*}{\rotcell{M1.A3}} & \multirow{6}{*}{\rotcell{M1.B1}} & \multirow{2}{*}{50}  & Doubly   & 0.944   & 0      & 0.948   & 0.002   & 0      & 0       & 0.960   & 0      & 0.956   & 0.498   & 0      & 0       & 0.950   & 0.392  & 0.960   & 0.850   & 0.008  & 0.008    \\
			&                                  &                      & Mixed    & 0.950   & 0      & 0.930   & 0       & 0      & 0       & 0.954   & 0      & 0.950   & 0.426   & 0      & 0       & 0.964   & 0.252  & 0.948   & 0.848   & 0      & 0        \\
			&                                  & \multirow{2}{*}{100} & Doubly   & 0.950   & 0      & 0.954   & 0       & 0      & 0       & 0.952   & 0      & 0.960   & 0.092   & 0      & 0       & 0.948   & 0.004  & 0.958   & 0.740   & 0      & 0        \\
			&                                  &                      & Mixed    & 0.952   & 0      & 0.958   & 0       & 0      & 0       & 0.946   & 0      & 0.962   & 0.058   & 0      & 0       & 0.956   & 0.006  & 0.954   & 0.736   & 0      & 0        \\ 
			\cline{2-22}
			& \multirow{6}{*}{\rotcell{M1.B2}} & \multirow{2}{*}{50}  & Doubly   & 0.956   & 0      & 0       & 0       & 0      & 0       & 0.960   & 0      & 0.022   & 0.518   & 0      & 0       & 0.940   & 0.372  & 0.620   & 0.836   & 0      & 0        \\
			&                                  &                      & Mixed    & 0.956   & 0      & 0       & 0.002   & 0      & 0       & 0.956   & 0      & 0       & 0.422   & 0      & 0       & 0.964   & 0.244  & 0.428   & 0.870   & 0      & 0        \\
			&                                  & \multirow{2}{*}{100} & Doubly   & 0.954   & 0      & 0       & 0       & 0      & 0       & 0.956   & 0      & 0       & 0.062   & 0      & 0       & 0.950   & 0.016  & 0.110   & 0.732   & 0      & 0        \\
			&                                  &                      & Mixed    & 0.948   & 0      & 0       & 0       & 0      & 0       & 0.960   & 0      & 0       & 0.050   & 0      & 0       & 0.956   & 0.008  & 0.034   & 0.728   & 0      & 0        \\ 
			\cline{2-22}
			& \multirow{6}{*}{\rotcell{M1.B3}} & \multirow{2}{*}{50}  & Doubly   & 0.934   & 0      & 0.006   & 0       & 0      & 0       & 0.956   & 0      & 0.632   & 0.492   & 0      & 0       & 0.952   & 0.340  & 0.896   & 0.874   & 0      & 0        \\
			&                                  &                      & Mixed    & 0.954   & 0      & 0.002   & 0       & 0      & 0       & 0.950   & 0      & 0.444   & 0.392   & 0      & 0       & 0.974   & 0.260  & 0.888   & 0.860   & 0      & 0        \\
			&                                  & \multirow{2}{*}{100} & Doubly   & 0.940   & 0      & 0       & 0       & 0      & 0       & 0.944   & 0      & 0.126   & 0.080   & 0      & 0       & 0.964   & 0.010  & 0.762   & 0.740   & 0      & 0        \\
			&                                  &                      & Mixed    & 0.956   & 0      & 0       & 0       & 0      & 0       & 0.942   & 0      & 0.066   & 0.058   & 0      & 0       & 0.958   & 0.006  & 0.714   & 0.674   & 0      & 0        \\
			\hline
		\end{tabular}
	}
\end{sidewaystable*}

\subsection{Second scenario (M2)}

This second study (M2) is motivated to analyse the quality of the proposed methods when others settings for error functions are used. In particular, it is assumed that $\epsilon_{ijk}(t)=20^{-1}B(t)$, where $B(t)$ is a standard brownian process with dispersion parameter $\sigma_\epsilon$. M2 has been inspired by \cite{Martinez2011}. The form for the functions of treatments, groups and interactions are the same than in M1. In Table \ref{tablaM2} appears the achieved results. The conclusions done above about the treatments are maintained.  However, the outcomes for the difference among groups and about the interactions are much better than in M1 when the corresponding $H_0$ is false. The error rate for $\sigma_\epsilon=0.20$ does not exceed the 8\% in any case. For its part, when jointly $\sigma_\epsilon=0.40$ and $n=50$ (because if $n=100$ the behaviour of the tests is really good) the acceptance proportion varies between 0.184-0.254 and 0.292-0.358 in MMM and DMM, respectively, for the case of parallelism. Likewise, for the effect of the groups, the proportion changes between 0.244-0.280 and 0.440-0.494 in MMM and DMM, respectively, when M2.B3 was applied. Therefore, the type of error is another key point in this kind of analysis.

	\begin{sidewaystable*}
	\centering
	\caption{Observed acceptance proportions for each scenario at a significance level 0.05 in the case of  simulation study M2.}\label{tablaM2}
	\resizebox{23cm}{!} {
		\begin{tabular}{|c|c|c|c|ccc|ccc|ccc|ccc|ccc|ccc|} 
			\cline{5-22}
			\multicolumn{1}{c}{}              & \multicolumn{1}{c}{}             & \multicolumn{1}{c}{} &          & \multicolumn{6}{c|}{$\sigma_\epsilon$=0.10}                & \multicolumn{6}{c|}{$\sigma_\epsilon$=0.20}                & \multicolumn{6}{c|}{$\sigma_\epsilon$=0.40}                 \\ 
			\cline{5-22}
			\multicolumn{1}{c}{}              & \multicolumn{1}{c}{}             & \multicolumn{1}{c}{} &          & \multicolumn{3}{c|}{M2.I1} & \multicolumn{3}{c|}{M2.I2} & \multicolumn{3}{c|}{M2.I1} & \multicolumn{3}{c|}{M2.I2} & \multicolumn{3}{c|}{M2.I1} & \multicolumn{3}{c|}{M2.I2}  \\ 
			\cline{3-22}
			\multicolumn{1}{c}{}              &                                  & n                    & Approach & Parall. & Treat. & Group   & Parall. & Treat. & Group   & Parall. & Treat. & Group   & Parall. & Treat. & Group   & Parall. & Treat. & Group   & Parall. & Treat. & Group    \\ 
			\hline
			\multirow{12}{*}{\rotcell{M2.A1}} & \multirow{6}{*}{\rotcell{M2.B1}} & \multirow{2}{*}{50}  & Doubly   & 0.968   & 0.960  & 0.968   & 0       & 0      & 0       & 0.948   & 0.950  & 0.944   & 0.028   & 0      & 0       & 0.956   & 0.940  & 0.948   & 0.342   & 0      & 0        \\
			&                                  &                      & Mixed    & 0.962   & 0.974  & 0.958   & 0       & 0      & 0       & 0.962   & 0.972  & 0.954   & 0.006   & 0      & 0       & 0.950   & 0.964  & 0.946   & 0.220   & 0      & 0        \\
			&                                  & \multirow{2}{*}{100} & Doubly   & 0.946   & 0.942  & 0.956   & 0       & 0      & 0       & 0.946   & 0.954  & 0.954   & 0       & 0      & 0       & 0.940   & 0.955  & 0.948   & 0.002   & 0      & 0        \\
			&                                  &                      & Mixed    & 0.954   & 0.958  & 0.930   & 0       & 0      & 0       & 0.942   & 0.946  & 0.924   & 0       & 0      & 0       & 0.950   & 0.968  & 0.944   & 0.004   & 0      & 0        \\ 
			\cline{2-22}
			& \multirow{6}{*}{\rotcell{M2.B2}} & \multirow{2}{*}{50}  & Doubly   & 0.940   & 0.952  & 0       & 0       & 0      & 0       & 0.958   & 0.948  & 0       & 0.016   & 0      & 0       & 0.954   & 0.960  & 0       & 0.320   & 0      & 0        \\
			&                                  &                      & Mixed    & 0.936   & 0.936  & 0       & 0       & 0      & 0       & 0.952   & 0.962  & 0       & 0       & 0      & 0       & 0.960   & 0.960  & 0       & 0.254   & 0      & 0        \\
			&                                  & \multirow{2}{*}{100} & Doubly   & 0.952   & 0.944  & 0       & 0       & 0      & 0       & 0.934   & 0.944  & 0       & 0       & 0      & 0       & 0.955   & 0.946  & 0       & 0.004   & 0      & 0        \\
			&                                  &                      & Mixed    & 0.950   & 0.962  & 0       & 0       & 0      & 0       & 0.942   & 0.928  & 0       & 0       & 0      & 0       & 0.950   & 0.956  & 0       & 0.006   & 0      & 0        \\ 
			\cline{2-22}
			& \multirow{6}{*}{\rotcell{M2.B3}} & \multirow{2}{*}{50}  & Doubly   & 0.952   & 0.938  & 0       & 0       & 0      & 0       & 0.954   & 0.948  & 0.078   & 0.024   & 0      & 0       & 0.950   & 0.938  & 0.494   & 0.298   & 0      & 0        \\
			&                                  &                      & Mixed    & 0.942   & 0.962  & 0       & 0       & 0      & 0       & 0.950   & 0.960  & 0.008   & 0.004   & 0      & 0       & 0.948   & 0.950  & 0.280   & 0.246   & 0      & 0        \\
			&                                  & \multirow{2}{*}{100} & Doubly   & 0.954   & 0.948  & 0       & 0       & 0      & 0       & 0.964   & 0.950  & 0       & 0       & 0      & 0       & 0.954   & 0.942  & 0.036   & 0.014   & 0      & 0        \\
			&                                  &                      & Mixed    & 0.964   & 0.958  & 0       & 0       & 0      & 0       & 0.958   & 0.946  & 0       & 0       & 0      & 0       & 0.962   & 0.946  & 0       & 0.004   & 0      & 0        \\ 
			\hline
			\multirow{12}{*}{\rotcell{M2.A2}} & \multirow{6}{*}{\rotcell{M2.B1}} & \multirow{2}{*}{50}  & Doubly   & 0.944   & 0      & 0.952   & 0       & 0      & 0       & 0.944   & 0      & 0.956   & 0.022   & 0      & 0       & 0.944   & 0      & 0.944   & 0.314   & 0      & 0        \\
			&                                  &                      & Mixed    & 0.948   & 0      & 0.938   & 0       & 0      & 0       & 0.942   & 0      & 0.920   & 0.008   & 0      & 0       & 0.942   & 0      & 0.932   & 0.184   & 0      & 0        \\
			&                                  & \multirow{2}{*}{100} & Doubly   & 0.948   & 0      & 0.952   & 0       & 0      & 0       & 0.970   & 0      & 0.954   & 0       & 0      & 0       & 0.936   & 0      & 0.940   & 0.006   & 0      & 0        \\
			&                                  &                      & Mixed    & 0.942   & 0      & 0.956   & 0       & 0      & 0       & 0.954   & 0      & 0.930   & 0       & 0      & 0       & 0.956   & 0      & 0.956   & 0.006   & 0      & 0        \\ 
			\cline{2-22}
			& \multirow{6}{*}{\rotcell{M2.B2}} & \multirow{2}{*}{50}  & Doubly   & 0.938   & 0      & 0       & 0       & 0      & 0       & 0.968   & 0      & 0       & 0.028   & 0      & 0       & 0.946   & 0      & 0       & 0.316   & 0      & 0        \\
			&                                  &                      & Mixed    & 0.946   & 0      & 0       & 0       & 0      & 0       & 0.948   & 0      & 0       & 0.010   & 0      & 0       & 0.964   & 0      & 0       & 0.224   & 0      & 0        \\
			&                                  & \multirow{2}{*}{100} & Doubly   & 0.942   & 0      & 0       & 0       & 0      & 0       & 0.942   & 0      & 0       & 0       & 0      & 0       & 0.956   & 0      & 0       & 0.016   & 0      & 0        \\
			&                                  &                      & Mixed    & 0.952   & 0      & 0       & 0       & 0      & 0       & 0.960   & 0      & 0       & 0       & 0      & 0       & 0.954   & 0      & 0       & 0.006   & 0      & 0        \\ 
			\cline{2-22}
			& \multirow{6}{*}{\rotcell{M2.B3}} & \multirow{2}{*}{50}  & Doubly   & 0.944   & 0      & 0       & 0       & 0      & 0       & 0.956   & 0      & 0.080   & 0.020   & 0      & 0       & 0.960   & 0      & 0.440   & 0.292   & 0      & 0        \\
			&                                  &                      & Mixed    & 0.954   & 0      & 0       & 0       & 0      & 0       & 0.942   & 0      & 0.014   & 0.010   & 0      & 0       & 0.948   & 0      & 0.278   & 0.200   & 0      & 0        \\
			&                                  & \multirow{2}{*}{100} & Doubly   & 0.940   & 0      & 0       & 0       & 0      & 0       & 0.956   & 0      & 0       & 0       & 0      & 0       & 0.930   & 0      & 0.012   & 0.010   & 0      & 0        \\
			&                                  &                      & Mixed    & 0.960   & 0      & 0       & 0       & 0      & 0       & 0.944   & 0      & 0       & 0       & 0      & 0       & 0.962   & 0      & 0.008   & 0.002   & 0      & 0        \\ 
			\hline
			\multirow{12}{*}{\rotcell{M2.A3}} & \multirow{6}{*}{\rotcell{M2.B1}} & \multirow{2}{*}{50}  & Doubly   & 0.946   & 0      & 0.934   & 0       & 0      & 0       & 0.956   & 0.052  & 0.950   & 0.032   & 0      & 0       & 0.930   & 0.328  & 0.938   & 0.316   & 0      & 0        \\
			&                                  &                      & Mixed    & 0.954   & 0      & 0.962   & 0       & 0      & 0       & 0.962   & 0.006  & 0.948   & 0.008   & 0      & 0       & 0.966   & 0.270  & 0.952   & 0.230   & 0      & 0        \\
			&                                  & \multirow{2}{*}{100} & Doubly   & 0.954   & 0      & 0.956   & 0       & 0      & 0       & 0.962   & 0      & 0.974   & 0       & 0      & 0       & 0.936   & 0.010  & 0.928   & 0.006   & 0      & 0        \\
			&                                  &                      & Mixed    & 0.960   & 0      & 0.960   & 0       & 0      & 0       & 0.950   & 0      & 0.948   & 0       & 0      & 0       & 0.960   & 0.012  & 0.958   & 0.008   & 0      & 0        \\ 
			\cline{2-22}
			& \multirow{6}{*}{\rotcell{M2.B2}} & \multirow{2}{*}{50}  & Doubly   & 0.954   & 0      & 0       & 0       & 0      & 0       & 0.942   & 0.050  & 0       & 0.020   & 0      & 0       & 0.948   & 0.408  & 0       & 0.308   & 0      & 0        \\
			&                                  &                      & Mixed    & 0.966   & 0      & 0       & 0       & 0      & 0       & 0.948   & 0.018  & 0       & 0.006   & 0      & 0       & 0.950   & 0.282  & 0       & 0.230   & 0      & 0        \\
			&                                  & \multirow{2}{*}{100} & Doubly   & 0.958   & 0      & 0       & 0       & 0      & 0       & 0.954   & 0      & 0       & 0       & 0      & 0       & 0.976   & 0.012  & 0       & 0.012   & 0      & 0        \\
			&                                  &                      & Mixed    & 0.930   & 0      & 0       & 0       & 0      & 0       & 0.926   & 0      & 0       & 0       & 0      & 0       & 0.948   & 0.009  & 0       & 0.006   & 0      & 0        \\ 
			\cline{2-22}
			& \multirow{6}{*}{\rotcell{M2.B3}} & \multirow{2}{*}{50}  & Doubly   & 0.946   & 0      & 0       & 0       & 0      & 0       & 0.964   & 0.052  & 0.064   & 0.034   & 0      & 0       & 0.958   & 0.394  & 0.464   & 0.358   & 0      & 0        \\
			&                                  &                      & Mixed    & 0.970   & 0      & 0       & 0       & 0      & 0       & 0.968   & 0.010  & 0.010   & 0.008   & 0      & 0       & 0.952   & 0.254  & 0.244   & 0.206   & 0      & 0        \\
			&                                  & \multirow{2}{*}{100} & Doubly   & 0.954   & 0      & 0       & 0       & 0      & 0       & 0.948   & 0      & 0       & 0       & 0      & 0       & 0.948   & 0.014  & 0.030   & 0.012   & 0      & 0        \\
			&                                  &                      & Mixed    & 0.940   & 0      & 0       & 0       & 0      & 0       & 0.944   & 0      & 0       & 0       & 0      & 0       & 0.936   & 0.004  & 0.008   & 0       & 0      & 0        \\
			\hline
		\end{tabular}
	}
\end{sidewaystable*}

\subsection{Third scenario (M3)}
M3 is carried out under the same conditions than in M1 except the form of the interaction functions which were modified. The reason of contemplating this scenario is due to the fact that the obtained results for testing the hypothesis of parallelism were unsatisfactory in M1. Hence, with this scenario it is intended to study the impact of changing the form of the functions in the power of the tests. Consequently, the particular forms for interaction functions are the following

\begin{enumerate}
	\item[] M3.I1: $\theta_{ij}(t)=sin(\pi t)^{13}, \ i=1,2,3;\ j=1,2;$		
	\item[] M3.I2: $\theta_{ij}(t)=sin(\pi t)^{21-2ij},\ i=1,2,3;\  j=1,2.$
\end{enumerate}

Table \ref{tablaM3} contains the outcomes of this study. The conclusions about the hypothesis of the treatment and group effects are similar to that given in the first study. Nevertheless, we notice an important improvement in relation to the hypothesis of parallelism. In particular, when null hypothesis is false and $\sigma_\epsilon=0.40$, the error rate is lower than 9.2\% for $n=50$ and being 0\% for $n=100$. Thus, this study lay bare another interesting factor that influences in the performance of the tests, i.e., the quality of the test relies on the shapes considered for the curves.

\begin{sidewaystable*}
	\centering
	\caption{Observed acceptance proportions for each scenario at a significance level 0.05 in the case of simulation study M3.}\label{tablaM3}
	\resizebox{23cm}{!} {
		\begin{tabular}{|c|c|c|c|ccc|ccc|ccc|ccc|ccc|ccc|} 
			\cline{5-22}
			\multicolumn{1}{c}{}              & \multicolumn{1}{c}{}             & \multicolumn{1}{c}{} &          & \multicolumn{6}{c|}{$\sigma_\epsilon$=0.10}                & \multicolumn{6}{c|}{$\sigma_\epsilon$=0.20}                & \multicolumn{6}{c|}{$\sigma_\epsilon$=0.40}                 \\ 
			\cline{5-22}
			\multicolumn{1}{c}{}              & \multicolumn{1}{c}{}             & \multicolumn{1}{c}{} &          & \multicolumn{3}{c|}{M3.I1} & \multicolumn{3}{c|}{M3.I2} & \multicolumn{3}{c|}{M3.I1} & \multicolumn{3}{c|}{M3.I2} & \multicolumn{3}{c|}{M3.I1} & \multicolumn{3}{c|}{M3.I2}  \\ 
			\cline{3-22}
			\multicolumn{1}{c}{}              &                                  & n                    & Approach & Parall. & Treat. & Group   & Parall. & Treat. & Group   & Parall. & Treat. & Group   & Parall. & Treat. & Group   & Parall. & Treat. & Group   & Parall. & Treat. & Group    \\ 
			\hline
			\multirow{12}{*}{\rotcell{M3.A1}} & \multirow{6}{*}{\rotcell{M3.B1}} & \multirow{2}{*}{50}  & Doubly   & 0.950   & 0.954  & 0.944   & 0       & 0      & 0       & 0.950   & 0.932  & 0.946   & 0       & 0      & 0       & 0.954   & 0.930  & 0.970   & 0.074   & 0      & 0.008    \\
			&                                  &                      & Mixed    & 0.940   & 0.952  & 0.930   & 0       & 0      & 0       & 0.944   & 0.950  & 0.930   & 0       & 0      & 0       & 0.948   & 0.944  & 0.93    & 0.038   & 0      & 0.016    \\
			&                                  & \multirow{2}{*}{100} & Doubly   & 0.944   & 0.938  & 0.950   & 0       & 0      & 0       & 0.960   & 0.954  & 0.948   & 0       & 0      & 0       & 0.950   & 0.960  & 0.950   & 0       & 0      & 0        \\
			&                                  &                      & Mixed    & 0.946   & 0.958  & 0.938   & 0       & 0      & 0       & 0.948   & 0.950  & 0.934   & 0       & 0      & 0       & 0.952   & 0.954  & 0.970   & 0       & 0      & 0        \\ 
			\cline{2-22}
			& \multirow{6}{*}{\rotcell{M3.B2}} & \multirow{2}{*}{50}  & Doubly   & 0.956   & 0.952  & 0       & 0       & 0      & 0       & 0.938   & 0.940  & 0.012   & 0       & 0      & 0       & 0.936   & 0.944  & 0.592   & 0.092   & 0      & 0        \\
			&                                  &                      & Mixed    & 0.956   & 0.960  & 0       & 0       & 0      & 0       & 0.956   & 0.960  & 0       & 0       & 0      & 0       & 0.950   & 0.946  & 0.390   & 0.046   & 0      & 0        \\
			&                                  & \multirow{2}{*}{100} & Doubly   & 0.956   & 0.956  & 0       & 0       & 0      & 0       & 0.954   & 0.940  & 0       & 0       & 0      & 0       & 0.952   & 0.954  & 0.140   & 0       & 0      & 0        \\
			&                                  &                      & Mixed    & 0.964   & 0.962  & 0       & 0       & 0      & 0       & 0.948   & 0.956  & 0       & 0       & 0      & 0       & 0.946   & 0.950  & 0.056   & 0       & 0      & 0        \\ 
			\cline{2-22}
			& \multirow{6}{*}{\rotcell{M3.B3}} & \multirow{2}{*}{50}  & Doubly   & 0.944   & 0.936  & 0.004   & 0       & 0      & 0       & 0.946   & 0.956  & 0.652   & 0       & 0      & 0       & 0.958   & 0.940  & 0.890   & 0.066   & 0      & 0        \\
			&                                  &                      & Mixed    & 0.954   & 0.946  & 0       & 0       & 0      & 0       & 0.938   & 0.930  & 0.418   & 0       & 0      & 0       & 0.942   & 0.950  & 0.854   & 0.018   & 0      & 0        \\
			&                                  & \multirow{2}{*}{100} & Doubly   & 0.940   & 0.946  & 0       & 0       & 0      & 0       & 0.934   & 0.942  & 0.160   & 0       & 0      & 0       & 0.942   & 0.938  & 0.752   & 0       & 0      & 0        \\
			&                                  &                      & Mixed    & 0.952   & 0.954  & 0       & 0       & 0      & 0       & 0.934   & 0.952  & 0.062   & 0       & 0      & 0       & 0.950   & 0.940  & 0.720   & 0       & 0      & 0        \\ 
			\hline
			\multirow{12}{*}{\rotcell{M3.A2}} & \multirow{6}{*}{\rotcell{M3.B1}} & \multirow{2}{*}{50}  & Doubly   & 0.970   & 0      & 0.954   & 0       & 0      & 0       & 0.958   & 0      & 0.964   & 0       & 0      & 0       & 0.962   & 0      & 0.944   & 0.086   & 0      & 0.010    \\
			&                                  &                      & Mixed    & 0.930   & 0      & 0.934   & 0       & 0      & 0       & 0.940   & 0      & 0.958   & 0       & 0      & 0       & 0.948   & 0      & 0.932   & 0.038   & 0      & 0.008    \\
			&                                  & \multirow{2}{*}{100} & Doubly   & 0.958   & 0      & 0.960   & 0       & 0      & 0       & 0.948   & 0      & 0.946   & 0       & 0      & 0       & 0.950   & 0      & 0.940   & 0       & 0      & 0        \\
			&                                  &                      & Mixed    & 0.954   & 0      & 0.950   & 0       & 0      & 0       & 0.946   & 0      & 0.930   & 0       & 0      & 0       & 0.964   & 0      & 0.946   & 0       & 0      & 0        \\ 
			\cline{2-22}
			& \multirow{6}{*}{\rotcell{M3.B2}} & \multirow{2}{*}{50}  & Doubly   & 0.938   & 0      & 0       & 0       & 0      & 0       & 0.954   & 0      & 0.006   & 0       & 0      & 0       & 0.948   & 0      & 0.618   & 0.076   & 0      & 0        \\
			&                                  &                      & Mixed    & 0.964   & 0      & 0       & 0       & 0      & 0       & 0.956   & 0      & 0       & 0       & 0      & 0       & 0.956   & 0      & 0.426   & 0.038   & 0      & 0.002    \\
			&                                  & \multirow{2}{*}{100} & Doubly   & 0.960   & 0      & 0       & 0       & 0      & 0       & 0.940   & 0      & 0       & 0       & 0      & 0       & 0.942   & 0      & 0.122   & 0       & 0      & 0        \\
			&                                  &                      & Mixed    & 0.966   & 0      & 0       & 0       & 0      & 0       & 0.946   & 0      & 0       & 0       & 0      & 0       & 0.950   & 0      & 0.050   & 0       & 0      & 0        \\ 
			\cline{2-22}
			& \multirow{6}{*}{\rotcell{M3.B3}} & \multirow{2}{*}{50}  & Doubly   & 0.944   & 0      & 0.020   & 0       & 0      & 0       & 0.958   & 0      & 0.638   & 0       & 0      & 0       & 0.956   & 0      & 0.880   & 0.086   & 0      & 0        \\
			&                                  &                      & Mixed    & 0.942   & 0      & 0       & 0       & 0      & 0       & 0.942   & 0      & 0.468   & 0       & 0      & 0       & 0.930   & 0      & 0.852   & 0.036   & 0      & 0.002    \\
			&                                  & \multirow{2}{*}{100} & Doubly   & 0.934   & 0      & 0       & 0       & 0      & 0       & 0.942   & 0      & 0.130   & 0       & 0      & 0       & 0.956   & 0      & 0.790   & 0       & 0      & 0        \\
			&                                  &                      & Mixed    & 0.962   & 0      & 0       & 0       & 0      & 0       & 0.952   & 0      & 0.078   & 0       & 0      & 0       & 0.950   & 0      & 0.714   & 0       & 0      & 0        \\ 
			\hline
			\multirow{12}{*}{\rotcell{M3.A3}} & \multirow{6}{*}{\rotcell{M3.B1}} & \multirow{2}{*}{50}  & Doubly   & 0.954   & 0      & 0.960   & 0       & 0      & 0       & 0.952   & 0      & 0.952   & 0       & 0      & 0       & 0.954   & 0.326  & 0.942   & 0.086   & 0      & 0.008    \\
			&                                  &                      & Mixed    & 0.958   & 0      & 0.946   & 0       & 0      & 0       & 0.954   & 0      & 0.932   & 0       & 0      & 0       & 0.946   & 0.202  & 0.932   & 0.026   & 0      & 0.006    \\
			&                                  & \multirow{2}{*}{100} & Doubly   & 0.968   & 0      & 0.964   & 0       & 0      & 0       & 0.950   & 0      & 0.956   & 0       & 0      & 0       & 0.966   & 0.010  & 0.946   & 0       & 0      & 0        \\
			&                                  &                      & Mixed    & 0.942   & 0      & 0.934   & 0       & 0      & 0       & 0.964   & 0      & 0.970   & 0       & 0      & 0       & 0.946   & 0.002  & 0.952   & 0       & 0      & 0        \\ 
			\cline{2-22}
			& \multirow{6}{*}{\rotcell{M3.B2}} & \multirow{2}{*}{50}  & Doubly   & 0.948   & 0      & 0       & 0       & 0      & 0       & 0.948   & 0      & 0.014   & 0       & 0      & 0       & 0.942   & 0.354  & 0.622   & 0.076   & 0      & 0        \\
			&                                  &                      & Mixed    & 0.964   & 0      & 0       & 0       & 0      & 0       & 0.930   & 0      & 0       & 0       & 0      & 0       & 0.962   & 0.248  & 0.410   & 0.042   & 0      & 0        \\
			&                                  & \multirow{2}{*}{100} & Doubly   & 0.950   & 0      & 0       & 0       & 0      & 0       & 0.940   & 0      & 0       & 0       & 0      & 0       & 0.934   & 0.014  & 0.134   & 0       & 0      & 0        \\
			&                                  &                      & Mixed    & 0.960   & 0      & 0       & 0       & 0      & 0       & 0.940   & 0      & 0       & 0       & 0      & 0       & 0.954   & 0.004  & 0.064   & 0       & 0      & 0        \\ 
			\cline{2-22}
			& \multirow{6}{*}{\rotcell{M3.B3}} & \multirow{2}{*}{50}  & Doubly   & 0.946   & 0      & 0.002   & 0       & 0      & 0       & 0.962   & 0      & 0.648   & 0       & 0      & 0       & 0.958   & 0.338  & 0.912   & 0.084   & 0      & 0        \\
			&                                  &                      & Mixed    & 0.960   & 0      & 0       & 0       & 0      & 0       & 0.946   & 0      & 0.444   & 0       & 0      & 0       & 0.960   & 0.240  & 0.870   & 0.026   & 0      & 0        \\
			&                                  & \multirow{2}{*}{100} & Doubly   & 0.960   & 0      & 0       & 0       & 0      & 0       & 0.932   & 0      & 0.124   & 0       & 0      & 0       & 0.948   & 0.004  & 0.800   & 0       & 0      & 0        \\
			&                                  &                      & Mixed    & 0.960   & 0      & 0       & 0       & 0      & 0       & 0.960   & 0      & 0.070   & 0       & 0      & 0       & 0.944   & 0.002  & 0.706   & 0       & 0      & 0        \\
			\hline
		\end{tabular}
	}
\end{sidewaystable*}

\subsection{Fourth scenario (M4)}
This study is worked out to corroborate the last affirmation made in previous section.  For that purpose, M4 presents the same characteristics than M1 but interchanging the functions of the groups and interactions, that is, now we have
\begin{enumerate}
	\item[] M4.B1: $\beta_j(t)=[sin(2\pi t^2)]^5, j=1,2;$		
	\item[] M4.B2: $\beta_j(t)=[sin(2\pi t^2)]^{3+2j},\ j=1,2;$
	\item[] M4.B3: $\beta_j(t)=[sin(2\pi t^2)]^{5+2j},\  j=1,2;$
	\item[] M4.I1: $\theta_{ij}(t)=0.05\times |sin(4\pi t)|,\ i=1,2,3; \ j=1,2;$		
	\item[] M4.I2: $\theta_{ij}(t)=(0.025\times ij)\times |sin(4\pi t)|,\ i=1,2,3; \ j=1,2.$
\end{enumerate}

The obtained results in M4 (see Table \ref{tablaM4}) confirm the suspicions about the importance of the curves form in the performance of the tests. The investigation brings excellent outcomes until $\sigma_\epsilon=0.20$. It is only appreciated an error rate slightly high for testing the parallelism of the profiles in DMM when $n=50$ and M4.I2 are considered at the same time (as maximum, the acceptance proportions reaches 10.8\%). On the other hand, when $\sigma_\epsilon=0.40$ two different behaviours are detected:

\begin{enumerate}
	\item For all cases where M4.I1 is considered, the behaviour of the tests is really acceptable for the group effect (there are only little deviations when DMM is employed for $n=50$ y M4.B3). Regarding the treatments, the remarks are similar to the rest of the previously applied models. Not a single problem was discovered for testing the hypothesis about interactions. As it has already been commented, the power of the tests is actually satisfactory when $H_0$ is true for any effect.  
	\item For all cases where M4.I2 is considered, the results are deficient for checking the parallelism. Besides, there are some problems for testing differences among groups when $n=50$ and M4.B1 are assumed, being the first time that it happens during the simulation. This should not occur because, although M4.B1 represents the case of no difference among groups, it is considered that the interaction depends on each level of the treatments and groups.  
\end{enumerate}

To sum up, this exhaustive simulation study displays sufficient evidences for concluding that this new methodologies, based on basis expansion of sample curves, is a suitable candidate to deal with FANOVA-RM problem. We are only slightly concerned about the variability of the power of the tests when the dispersion parameter is great. However, it is important to keep in mind that the subject effect also plays a fundamental role in the analysis and here it is assumed a high value for $\sigma_k$. This fact produces important noise in the curves and the tests could convert into less conservative. Hence, this is another reason that endorses the goodness of the approaches presented in this paper, since even when the variability among subjects is large, the tests works very well in general terms.

Finally, although the assumptions are satisfied by the own construction of the models, the normality, variance homogeneity and the sphericity (for MMM) are checked with a ratio of rejection lower than 5\% of the cases. The developed simulation and the results shown in this section have been computationally implemented in R-cran. 

\begin{sidewaystable*}
	\centering
	\caption{Observed acceptance proportions for each scenario at a significance level 0.05 in the case of simulation study M4.}\label{tablaM4}
	\resizebox{23cm}{!} {
		\begin{tabular}{|c|c|c|c|ccc|ccc|ccc|ccc|ccc|ccc|} 
			\cline{5-22}
			\multicolumn{1}{c}{}              & \multicolumn{1}{c}{}             & \multicolumn{1}{c}{} &          & \multicolumn{6}{c|}{$\sigma_\epsilon$=0.10}                & \multicolumn{6}{c|}{$\sigma_\epsilon$=0.20}                & \multicolumn{6}{c|}{$\sigma_\epsilon$=0.40}                 \\ 
			\cline{5-22}
			\multicolumn{1}{c}{}              & \multicolumn{1}{c}{}             & \multicolumn{1}{c}{} &          & \multicolumn{3}{c|}{M4.I1} & \multicolumn{3}{c|}{M4.I2} & \multicolumn{3}{c|}{M4.I1} & \multicolumn{3}{c|}{M4.I2} & \multicolumn{3}{c|}{M4.I1} & \multicolumn{3}{c|}{M4.I2}  \\ 
			\cline{3-22}
			\multicolumn{1}{c}{}              &                                  & n                    & Approach & Parall. & Treat. & Group   & Parall. & Treat. & Group   & Parall. & Treat. & Group   & Parall. & Treat. & Group   & Parall. & Treat. & Group   & Parall. & Treat. & Group    \\ 
			\hline
			\multirow{12}{*}{\rotcell{M4.A1}} & \multirow{6}{*}{\rotcell{M4.B1}} & \multirow{2}{*}{50}  & Doubly   & 0.952   & 0.956  & 0.936   & 0       & 0      & 0       & 0.938   & 0.950  & 0.954   & 0.078   & 0      & 0       & 0.932   & 0.946  & 0.942   & 0.738   & 0.002  & 0.430    \\
			&                                  &                      & Mixed    & 0.960   & 0.942  & 0.952   & 0       & 0      & 0       & 0.948   & 0.954  & 0.936   & 0.068   & 0      & 0       & 0.966   & 0.960  & 0.960   & 0.712   & 0      & 0.332    \\
			&                                  & \multirow{2}{*}{100} & Doubly   & 0.940   & 0.944  & 0.940   & 0       & 0      & 0       & 0.940   & 0.952  & 0.950   & 0       & 0      & 0       & 0.950   & 0.930  & 0.944   & 0.374   & 0      & 0.016    \\
			&                                  &                      & Mixed    & 0.956   & 0.948  & 0.930   & 0       & 0      & 0       & 0.952   & 0.958  & 0.942   & 0       & 0      & 0       & 0.936   & 0.944  & 0.930   & 0.320   & 0      & 0.014    \\ 
			\cline{2-22}
			& \multirow{6}{*}{\rotcell{M4.B2}} & \multirow{2}{*}{50}  & Doubly   & 0.938   & 0.932  & 0       & 0       & 0      & 0       & 0.934   & 0.942  & 0       & 0.080   & 0      & 0       & 0.954   & 0.950  & 0       & 0.782   & 0.004  & 0        \\
			&                                  &                      & Mixed    & 0.950   & 0.956  & 0       & 0       & 0      & 0       & 0.956   & 0.946  & 0       & 0.050   & 0      & 0       & 0.946   & 0.946  & 0       & 0.722   & 0      & 0        \\
			&                                  & \multirow{2}{*}{100} & Doubly   & 0.940   & 0.940  & 0       & 0       & 0      & 0       & 0.940   & 0.942  & 0       & 0       & 0      & 0       & 0.944   & 0.954  & 0       & 0.364   & 0      & 0        \\
			&                                  &                      & Mixed    & 0.958   & 0.950  & 0       & 0       & 0      & 0       & 0.960   & 0.950  & 0       & 0       & 0      & 0       & 0.956   & 0.956  & 0       & 0.298   & 0      & 0        \\ 
			\cline{2-22}
			& \multirow{6}{*}{\rotcell{M4.B3}} & \multirow{2}{*}{50}  & Doubly   & 0.952   & 0.966  & 0       & 0       & 0      & 0       & 0.952   & 0.944  & 0       & 0.084   & 0      & 0       & 0.948   & 0.964  & 0.152   & 0.756   & 0      & 0        \\
			&                                  &                      & Mixed    & 0.944   & 0.942  & 0       & 0       & 0      & 0       & 0.948   & 0.934  & 0       & 0.048   & 0      & 0       & 0.948   & 0.944  & 0.026   & 0.694   & 0      & 0        \\
			&                                  & \multirow{2}{*}{100} & Doubly   & 0.948   & 0.952  & 0       & 0       & 0      & 0       & 0.954   & 0.954  & 0       & 0       & 0      & 0       & 0.924   & 0.946  & 0       & 0.356   & 0      & 0        \\
			&                                  &                      & Mixed    & 0.954   & 0.956  & 0       & 0       & 0      & 0       & 0.956   & 0.960  & 0       & 0       & 0      & 0       & 0.944   & 0.950  & 0       & 0.360   & 0      & 0        \\ 
			\hline
			\multirow{12}{*}{\rotcell{M4.A2}} & \multirow{6}{*}{\rotcell{M4.B1}} & \multirow{2}{*}{50}  & Doubly   & 0.954   & 0      & 0.954   & 0       & 0      & 0       & 0.942   & 0      & 0.944   & 0.086   & 0      & 0       & 0.942   & 0      & 0.952   & 0.746   & 0      & 0.436    \\
			&                                  &                      & Mixed    & 0.950   & 0      & 0.946   & 0       & 0      & 0       & 0.956   & 0      & 0.934   & 0.042   & 0      & 0       & 0.930   & 0      & 0.938   & 0.698   & 0      & 0.318    \\
			&                                  & \multirow{2}{*}{100} & Doubly   & 0.966   & 0      & 0.960   & 0       & 0      & 0       & 0.938   & 0      & 0.938   & 0       & 0      & 0       & 0.948   & 0      & 0.958   & 0.386   & 0      & 0.022    \\
			&                                  &                      & Mixed    & 0.946   & 0      & 0.942   & 0       & 0      & 0       & 0.946   & 0      & 0.936   & 0       & 0      & 0       & 0.964   & 0      & 0.938   & 0.338   & 0      & 0.018    \\ 
			\cline{2-22}
			& \multirow{6}{*}{\rotcell{M4.B2}} & \multirow{2}{*}{50}  & Doubly   & 0.954   & 0      & 0       & 0       & 0      & 0       & 0.948   & 0      & 0       & 0.106   & 0      & 0       & 0.934   & 0      & 0       & 0.766   & 0      & 0        \\
			&                                  &                      & Mixed    & 0.952   & 0      & 0       & 0       & 0      & 0       & 0.958   & 0      & 0       & 0.038   & 0      & 0       & 0.946   & 0      & 0       & 0.702   & 0      & 0        \\
			&                                  & \multirow{2}{*}{100} & Doubly   & 0.950   & 0      & 0       & 0       & 0      & 0       & 0.950   & 0      & 0       & 0       & 0      & 0       & 0.962   & 0      & 0       & 0.358   & 0      & 0        \\
			&                                  &                      & Mixed    & 0.948   & 0      & 0       & 0       & 0      & 0       & 0.958   & 0      & 0       & 0       & 0      & 0       & 0.938   & 0      & 0       & 0.330   & 0      & 0        \\ 
			\cline{2-22}
			& \multirow{6}{*}{\rotcell{M4.B3}} & \multirow{2}{*}{50}  & Doubly   & 0.944   & 0      & 0       & 0       & 0      & 0       & 0.964   & 0      & 0       & 0.090   & 0      & 0       & 0.954   & 0      & 0.132   & 0.776   & 0      & 0        \\
			&                                  &                      & Mixed    & 0.958   & 0      & 0       & 0       & 0      & 0       & 0.954   & 0      & 0       & 0.038   & 0      & 0       & 0.954   & 0      & 0.028   & 0.668   & 0      & 0        \\
			&                                  & \multirow{2}{*}{100} & Doubly   & 0.956   & 0      & 0       & 0       & 0      & 0       & 0.958   & 0      & 0       & 0       & 0      & 0       & 0.948   & 0      & 0       & 0.396   & 0      & 0        \\
			&                                  &                      & Mixed    & 0.958   & 0      & 0       & 0       & 0      & 0       & 0.972   & 0      & 0       & 0       & 0      & 0       & 0.952   & 0      & 0       & 0.344   & 0      & 0        \\ 
			\hline
			\multirow{12}{*}{\rotcell{M4.A3}} & \multirow{6}{*}{\rotcell{M4.B1}} & \multirow{2}{*}{50}  & Doubly   & 0.964   & 0      & 0.962   & 0       & 0      & 0       & 0.958   & 0      & 0.934   & 0.082   & 0      & 0       & 0.958   & 0.364  & 0.966   & 0.740   & 0.004  & 0.416    \\
			&                                  &                      & Mixed    & 0.946   & 0      & 0.954   & 0       & 0      & 0       & 0.954   & 0      & 0.930   & 0.042   & 0      & 0       & 0.944   & 0.246  & 0.946   & 0.700   & 0.002  & 0.304    \\
			&                                  & \multirow{2}{*}{100} & Doubly   & 0.960   & 0      & 0.952   & 0       & 0      & 0       & 0.950   & 0      & 0.964   & 0       & 0      & 0       & 0.958   & 0.014  & 0.960   & 0.40    & 0      & 0.018    \\
			&                                  &                      & Mixed    & 0.958   & 0      & 0.960   & 0       & 0      & 0       & 0.930   & 0      & 0.930   & 0       & 0      & 0       & 0.938   & 0      & 0.930   & 0.346   & 0      & 0.022    \\ 
			\cline{2-22}
			& \multirow{6}{*}{\rotcell{M4.B2}} & \multirow{2}{*}{50}  & Doubly   & 0.954   & 0      & 0       & 0       & 0      & 0       & 0.944   & 0      & 0       & 0.100   & 0      & 0       & 0.930   & 0.370  & 0       & 0.744   & 0.012  & 0        \\
			&                                  &                      & Mixed    & 0.956   & 0      & 0       & 0       & 0      & 0       & 0.956   & 0      & 0       & 0.046   & 0      & 0       & 0.944   & 0.214  & 0       & 0.656   & 0.002  & 0        \\
			&                                  & \multirow{2}{*}{100} & Doubly   & 0.952   & 0      & 0       & 0       & 0      & 0       & 0.946   & 0      & 0       & 0       & 0      & 0       & 0.954   & 0.016  & 0       & 0.386   & 0      & 0        \\
			&                                  &                      & Mixed    & 0.970   & 0      & 0       & 0       & 0      & 0       & 0.950   & 0      & 0       & 0       & 0      & 0       & 0.956   & 0.002  & 0       & 0.390   & 0      & 0        \\ 
			\cline{2-22}
			& \multirow{6}{*}{\rotcell{M4.B3}} & \multirow{2}{*}{50}  & Doubly   & 0.948   & 0      & 0       & 0       & 0      & 0       & 0.940   & 0      & 0       & 0.108   & 0      & 0       & 0.956   & 0.324  & 0.126   & 0.756   & 0.014  & 0        \\
			&                                  &                      & Mixed    & 0.930   & 0      & 0       & 0       & 0      & 0       & 0.950   & 0      & 0       & 0.068   & 0      & 0       & 0.944   & 0.248  & 0.016   & 0.682   & 0      & 0        \\
			&                                  & \multirow{2}{*}{100} & Doubly   & 0.950   & 0      & 0       & 0       & 0      & 0       & 0.948   & 0      & 0       & 0       & 0      & 0       & 0.954   & 0.006  & 0       & 0.404   & 0      & 0        \\
			&                                  &                      & Mixed    & 0.940   & 0      & 0       & 0       & 0      & 0       & 0.958   & 0      & 0       & 0       & 0      & 0       & 0.930   & 0.004  & 0       & 0.376   & 0      & 0        \\
			\hline
		\end{tabular}
	}
\end{sidewaystable*}

\section{Applications in Biomechanics}

In this section, two different applications with real data associated with human  body movement are developed.

\subsection{Human activity data}
\begin{figure}
	\centering
	\includegraphics[scale=0.48]{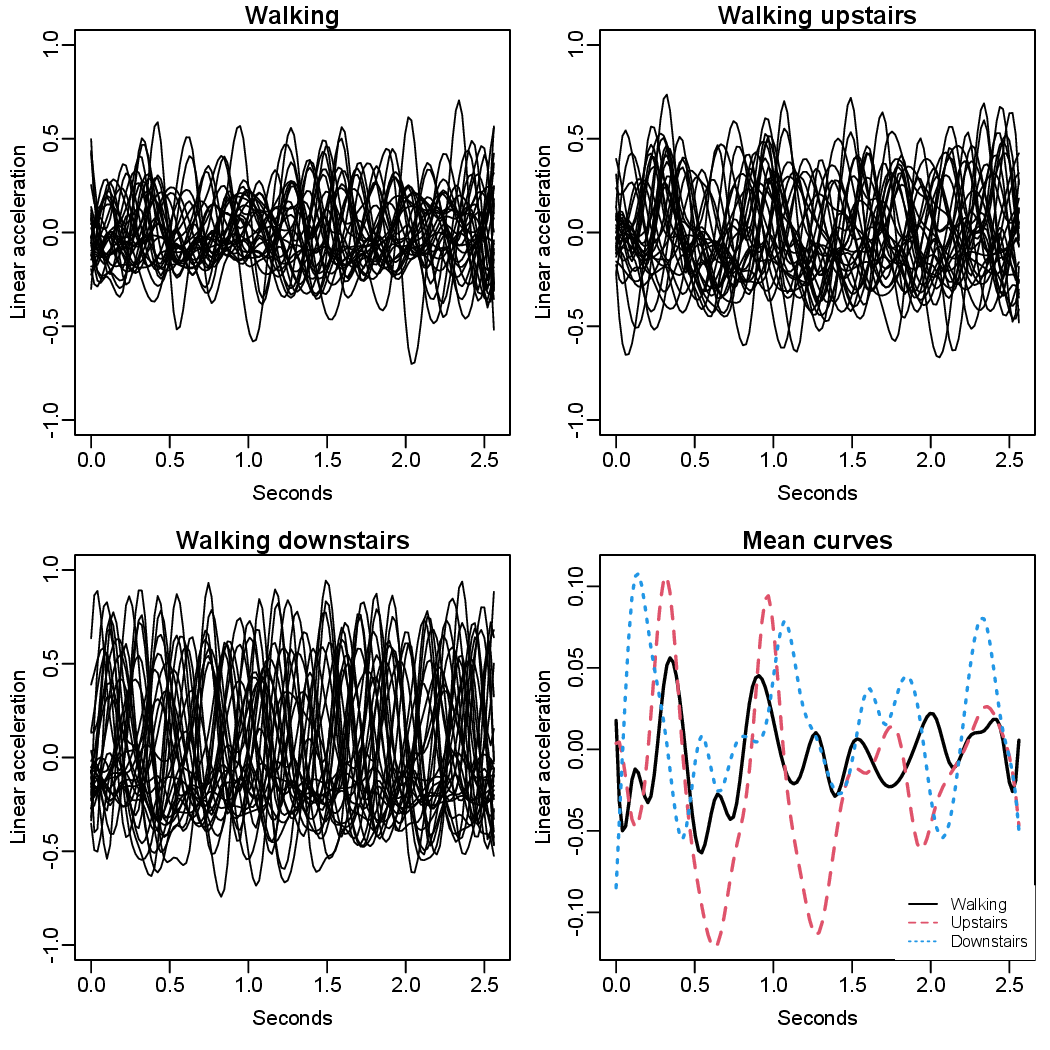}
	\caption{Sample group mean functions (bottom-right) and all the B-spline smoothed registered curves for each stimulus.}
	\label{curvesAP1}
\end{figure}

The data of this application correspond with a research carried out by \cite{Anguita2013} where the human activity recognition over 30 subjects was analysed. In the current manuscript, the study is focused on the variable called linear acceleration (metre per second squared) which is measured on the axis X. The information about this variable is recorded under three different treatments (walking, walking upstairs and walking downstairs) in 128 equidistant knots at the interval [0,2.56].  A subject was removed for being considered as outlier. This dataset was also used in \cite{Aguilera-Morillo20} for a functional linear discriminant analysis approach to classify a set of kinematic data. In the line of this work, sample curves were reconstructed through a cubic B-spline basis of dimension 27 with 25 equally spaced knots in the interval [0,2.56]. The smoothed curves for each treatment are displayed in Figure \ref{curvesAP1} together with the sample mean function of each group (bottom-right). Based off this graph, it seems reasonable to think that there are significant differences among the three stimulus, i.e., the linear acceleration on the axis X is affected by the type of movement. This fact is numerically corroborated by means of the FANOVA-RM approaches considered in this paper.

FANOVA-RM analysis conducted in this manuscript is summarised as follows. On the one hand, there is a single available group ($g=1$). Then, it is only possible to check whether there are differences between the treatments. On the other hand, DMM can not be considered because the sample space is smaller than the space of the variables ($n<pm$, being $n=29$, $p=27$ and $m=3$). Therefore, MMM is employed in order to test the differences aforementioned. Finally, the \textit{p-value} is calculated under the following conditions: 
\begin{enumerate}
	\item MMM is applied by assuming that the normality and the sphericity are satisfied. 
	\item Due to the fact that the multivariate sphericity assumption is rejected (the likelihood ratio test provided a \textit{p-value} lower than $\alpha=0.05$), an adjusted MMM by correcting the  degrees of freedom of F-statistic is performed. 
	\item Given that the normality assumption is also in question, the permutation testing procedure proposed by \cite{Gorecki15} is adapted for the repeated measures scenario. The steps of this procedure are described below:
	\begin{itemize}
		\item[(A)] Calculate the value of the test statistic $S_0$ for the original sample data.
		\item[(B)] For each subject, it is necessary to permute randomly without replacement its observed values on the treatments. If there were more than one group, once permuted the values of all subjects, the following step would be to join all subjects in 'a single group'. Later,  choose randomly without replacement $n_1$ observations for the first new group, then from the remainder of the observations draw randomly without replacement $n_2$ observations for the second new group, and repeat this process up to complete the $g$ groups. 
		\item[(C)] Compute the value of the test statistic for the new sample generated in previous step. 
		\item[(D)] Repeat steps (B)-(C) $F$ times, being $F$ a large number. Each achieved value  in (C) will be denoted by $S_f$ with $f=1,\ldots,F$. 
		\item[(E)]  Obtain the \textit{p-value} according next rule:
		$$
		p-value= \left\lbrace
		\begin{array}{l}
			\displaystyle \frac{1}{F}\sum_{f=1}^FI(S_f\leq S_0) \ \mathrm{for} \ S=W \\
			\\
			\displaystyle \frac{1}{F}\sum_{f=1}^FI(S_f\geq S_0) \ \mathrm{for}  \ S=LH,P,R \\
		\end{array}
		\right.
		$$
	\end{itemize}
\end{enumerate}

The \textit{p-values} obtained after applying the different methods described above figure in Table \ref{tablaAP1}. As a result, we can conclude that there are significant differences among the three types of stimulus on the linear acceleration on the axis X.    

\begin{table}
	\begin{center}
		\begin{tabular}{cccc}
			\hline 
			Stat. & MMM & Adjusted MMM & Permutation MMM \\ 
			\hline
			P & 0.0020 & 0.0010 & 0.0019 \\ 
			
			W & $<$0.0001 & 0.0001 & 0.0001 \\ 
			
			LH & $<$0.0001 & 0.0001 & 0.0009 \\ 
			
			R & $<$0.0001 & $<$0.0001 & 0.0001 \\ 
			\hline 
		\end{tabular} 
		\caption{P-values after applying MMM approach for FANOVA-RM with different scenarios.}\label{tablaAP1}
	\end{center}
\end{table}

\subsection{Dataset of running biomechanics}

\begin{figure}
	\centering
	\includegraphics[scale=0.48]{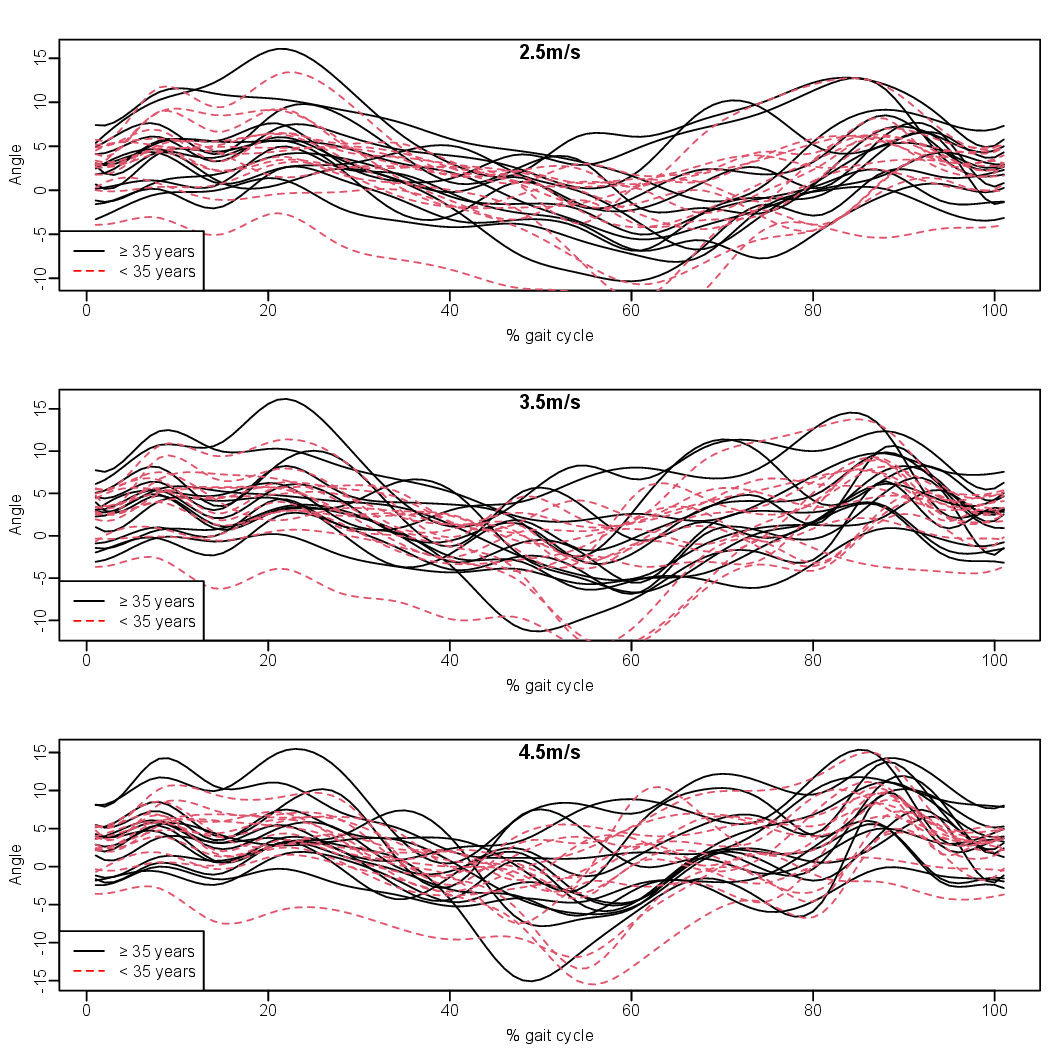}
	\caption{B-spline smoothed registered curves for each velocity according to the age group. }
	\label{curvesAP2}
\end{figure}

\begin{figure}
	\centering
	\includegraphics[scale=0.48]{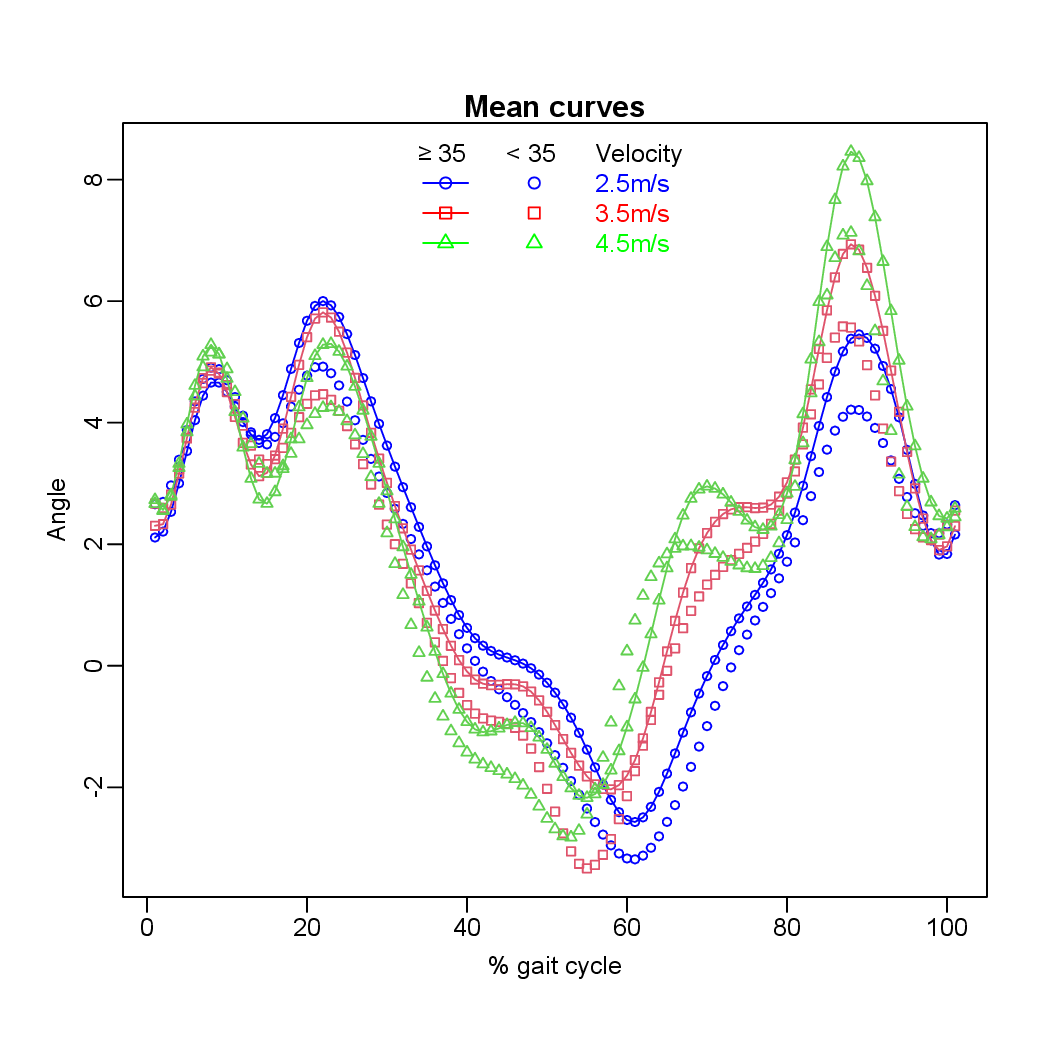}
	\caption{\textcolor{red}{Sample mean functions according to the age group and velocity.}}
	\label{mediasAP2}
\end{figure}

The public dataset available in \cite{Fukuchi2017} contains the biomechanical information of 28 regular runners. Concretely, lower-extremity kinematics and kinetics were registered meanwhile the subjects ran at different velocities (2.5 m/s, 3.5 m/s and 4.5 m/s) on an instrumental treadmill.  Other relevant data such as demographics information, running-training characteristics or previous injuries were also collected. The current application is focused on analysing the angle (in degrees) of right knee on the axis X, which has been recorded over 101 time points. In particular, the aim is to detect if there are significant differences in this functional variable among the different velocities (repeated measures) according to the age. The variable \textit{age} has been discretized in two independent groups. The first group is formed by 14 runners with less than 35 years old and the second one by the rest  ($\geq 35$ years). The functional reconstruction of the curves by means of a cubic B-spline basis of dimension 18 can be seen in Figure \ref{curvesAP2}. The mean curves of each age group under the different conditions are displayed in Figure \ref{mediasAP2}. We observe certain differences in the angle regarding the velocity, being greater for the velocity equals to 2m/s. Nevertheless, it is less clear whether the angle depends on the age group; the angle of the runners older than 35 year old is higher in all the domain, especially at the end of the cycle, but the discrepancies are not very noticeable. In order to validate statistically these assertions, the new methodology is applied.

In particular, this FANOVA-RM analysis contains the following characteristics: there are two independent groups of runners classified by the age, that is, $g=2$. The number of treatments is $m=3$, whereas $p=18$ because of the dimension of the functional reconstruction. In these conditions, only MMM can be considered. As the model assumptions are not verified, the permutation testing procedure aforementioned will be applied making use of the Pillai's trace statistic which is more robust than the other statistics in relation to the violation of model assumptions \citep{Olson1974}. The results of this analysis can be seen in Table \ref{tablaAP2}. 

\begin{table}
	\begin{center}
		\begin{tabular}{ccc}
			\hline
			Parallelism & Differences-groups & Differences-treatments \\
			\hline
			0.3852 & 0.7685 & 0.0020 \\
			\hline
		\end{tabular}
		\caption{P-values after applying the MMM approach for FANOVA-RM by means of permutation testing procedure.}\label{tablaAP2}
	\end{center}
\end{table}

The results show that the effect of the treatments does not depends on the levels of the age (there is not interaction). Additionally, it is corroborated that the differences between the age group are not significant, while the running speed plays an important role in the angle of the right knee on the axis X. 

\section{Conclusions}

Functional analysis of variance with repeated measures aims at checking if the mean functions of a functional response variable observed in different time periods or treatments are equal or not. In spite of its great interest in practice, only a few works related to this topic are available in the literature. The current manuscript is focused on addressing Two-Way FAVOVA-RM problem. The first factor represents the multiple levels in which the functional variable is observed (repeated measures), while the second one constitutes the independent groups in which the subjects of the sample are distributed (independent measures). Under this scenario, it is necessary to study both the between-group and intra-subject variability. As far as we know, this theoretical setting has not been ever dealt yet from a functional data analysis  viewpoint. Hence, a new approach based on basis expansion of sample curves is introduced in order to solve this problem. In particular, we prove that Two-Way FANOVA-RM model turns into Two-Way multivariate ANOVA-RM model for the multivariate response defined by the basis coefficients of the functional variable.  In this point, mixed multivariate model or doubly multivariate model can be conducted to take the intra-subject variability into account in the analysis. An extensive simulation study has shown that the multivariate mixed model approach has a better performance than the doubly multivariate model, although both approaches provide good results in general terms. Only in extreme situations (small differences among functions, small sample size or great dispersion) the  tests become conservative. The new methodology has also been applied to two real biomechanical datasets. In the first application, the objective is to evaluate how three type of stimulus affect  on the linear acceleration of human  movement, whereas the second study is focused on analysing the influence of age and speed in the knee flexion angle while running.


%
 \section*{Conflict of interest}
The authors declare that they have no conflict of interest.

\section*{Authors' contributions} 
All authors contributed equally to this work. All authors have read and agreed to the
published version of the manuscript.

\bibliographystyle{spbasic}      
\bibliography{Biblio_abb}   

%
%

\end{document}